\documentclass[aps,prd,nofootinbib,preprintnumbers,nofootinbibn,superscriptaddress]{revtex4}

\usepackage[english]{babel}				
\usepackage{mathtools}					
\usepackage{xcolor}						
\usepackage{amsmath}
\usepackage{amssymb}

\usepackage{graphicx}  
\usepackage{dcolumn}   
\usepackage{bm}        
\usepackage{amssymb}   
\usepackage[utf8]{inputenc}
\usepackage[english]{babel}
\usepackage{shapepar}

\usepackage{amsmath}
\usepackage{array}
\usepackage{fancybox}
\usepackage{multirow}
\usepackage{appendix}
\usepackage{epsfig}
\usepackage{cases}
\usepackage{forest}
\usepackage{stackrel}

\usepackage{graphicx}
\usepackage{dcolumn}
\usepackage{bm}
\usepackage{color}
\RequirePackage[colorlinks=true
,urlcolor=blue
,anchorcolor=blue
,citecolor=blue
,filecolor=blue
,linkcolor=blue
,menucolor=blue
,linktocpage=true
,pdfproducer=medialab
,pdfa=true
]{hyperref}

\newcommand\bs[1]{\boldsymbol{#1}}                      
\newcommand\dd{\mathrm{d}}                              
\newcommand\pp{\partial}                                
\newcommand\feq{\mathrel{\phantom{=}}}                  

\begin{document}


\title{Junction conditions in infinite derivative gravity}

\author{Ivan Kol\'a\v{r}}
\email{i.kolar@rug.nl}
\affiliation{Van Swinderen Institute, University of Groningen, 9747 AG, Groningen, The Netherlands}

\author{Francisco Jos\'{e} Maldonado Torralba}
\email{f.j.maldonado.torralba@rug.nl}
\affiliation{Van Swinderen Institute, University of Groningen, 9747 AG, Groningen, The Netherlands}
\affiliation{Cosmology and Gravity Group, Department of Mathematics and Applied Mathematics,
University of Cape Town, Rondebosch 7701, Cape Town, South Africa}

\author{Anupam Mazumdar}
\email{anupam.mazumdar@rug.nl}
\affiliation{Van Swinderen Institute, University of Groningen, 9747 AG, Groningen, The Netherlands}

\date{\today}

\begin{abstract}
The junction conditions for the infinite derivative gravity theory ${R{+}RF(\Box)R}$ are derived under the assumption that the conditions can be imposed by avoiding the `ill-defined expressions' in the theory of distributions term by term in infinite summations. We find that the junction conditions of such non-local theories are much more restrictive than in local theories, since the conditions comprise an infinite number of equations for the Ricci scalar. These conditions can constrain the geometry far beyond the matching hypersurface. Furthermore, we derive the junction field equations which are satisfied by the energy-momentum on the hypersurface. It turns out that the theory still allows some matter content on the hypersurface (without external flux and external tension), but with a traceless energy-momentum tensor. We also discuss the proper matching condition where no matter is concentrated on the hypersurface. Finally, we explore the possible applications and consequences of our results to the braneworld scenarios and star models. Particularly, we find that the internal tension is given purely by the trace of the energy-momentum tensor of the matter confined to the brane. Consequences of the junction conditions are illustrated on two simple examples of static and collapsing stars. It is demonstrated that even without solving the field equations the geometry on one side of the hypersurface can be determined to a great extent by the geometry on the other side if the Ricci scalar is analytic. We further show that some usual star models in the general relativity are no longer solutions of the infinite derivative gravity.
\end{abstract}

\maketitle


\section{Introduction}

In every physical theory, one must face the question of how to treat surfaces of discontinuity. For example, in electromagnetism, one can have charged surfaces that induce a `jump' on the value of the electric and magnetic fields \cite{Misner:1974qy}. The equations that relate the structure of this surface with discontinuities of the physical quantities are known as the \textit{junction conditions}. In a gravitational context, the situation is very similar, and the junction conditions can be derived for a hypersurface that divides the spacetime into two regions. This issue was first studied in the general relativity (GR) by Lanczos \cite{https://doi.org/10.1002/andp.19243791403}. Then, subsequent studies have formulated the junction conditions for timelike, spacelike, null, and general hypersurfaces, in terms of the intrinsic and extrinsic curvatures of such hypersurfaces \cite{darmois1927equations,Lichnerowicz:107002,o1952jump,bel1967conditions,taub1980space,bonnor1981junction,clarke1987junction,barrabes1989singular,Mars:1993mj}. Also, this junction conditions affect the field equations, allowing certain jumps in the energy-momentum tensor, first obtained for non-null hypersurfaces by Israel \cite{Israel:1966rt}, hence acquiring the name of \textit{Israel's junction conditions}. Null and general hypersurfaces were later studied in \cite{barrabes1991thin,Mars:1993mj}. 

Being such a crucial aspect of a theory, it is clear that the junction conditions are of immense importance in any modified theory of gravity. Examples of such studies include brane-world models \cite{Davis:2002gn,Battye:2001pb}, $f(R)$ gravity \cite{Deruelle:2007pt,Clifton:2012ry,Senovilla:2013vra}, Palatini $f(R)$ \cite{Olmo:2020fri}, $f(R)$ with torsion \cite{Vignolo:2018eco}, quadratic gravity \cite{Reina:2015gxa}, Einstein-Cartan theory \cite{Arkuszewski:1975fz}, metric-affine gravity \cite{Macias:2002sr}, generalised scalar-tensor theories \cite{Padilla:2012ze}, and extended teleparallel gravity \cite{delaCruz-Dombriz:2014zaa}.

In this paper, we will focus on a particular non-local theory usually known as the infinite derivative gravity (IDG) \cite{Biswas:2005qr,Biswas:2011ar}. For this theory, the non-local interaction arises due to the inclusion of operators with an infinite number of derivatives in the gravitational action.  The most general action has been built around Minkowski~\cite{Biswas:2011ar}, de Sitter and anti-de Sitter spaces~\cite{Biswas:2016etb}, and around cosmological bouncing background~\cite{Biswas:2005qr,Biswas:2010zk}. The graviton propagator of such theories can be modified to avoid any perturbative ghosts around a specific background. Also, the non-local gravitational interaction has been argued to improve UV aspects gravity at the quantum level~\cite{tomboulis1997superrenormalizable,Modesto:2011kw,Talaganis:2014ida,Abel:2019ufz,Abel:2019zou,Abel:2020gdi}. There were also several attempts in studying the initial value problem of IDG using diffusion equation method \cite{Calcagni:2018lyd} and constructing perturbative
Hamiltonian by introducing one extra time dimension \cite{Kolar:2020ezu}.

At the classical level, it has been shown that the linearized IDG can yield non-singular static solutions for point sources (including electromagnetic charges), p-branes (and cosmic strings), sources describing NUT charges \cite{Biswas:2011ar,Buoninfante:2018stt,Boos:2018bxf,Boos:2020ccj,Kolar:2020bpo}, spinning ring distributions \cite{Buoninfante:2018xif}. It was also found that the linearized IDG prevents mini-black-hole production for small masses \cite{Frolov:2015usa,Frolov:2015bia,Frolov:2015bta}. Exact solutions of the full IDG describing bouncing cosmologies, impulsive gravitational waves were obtained in~\cite{Biswas:2005qr,Kilicarslan:2018yxd,Dengiz:2020xbu}. An extension of IDG theories allowing a non-symmetric connection (retaining the absence of ghosts and non-singular linearized solutions) was proposed in \cite{delaCruz-Dombriz:2018aal,de2019ghost}.

The study of the junction conditions in non-local theories is of great interest. This is because one of the main aspects of the non-locality is that the fields interact directly over a greater distance, as the non-local scale modifies the UV aspects of gravitational interactions. Therefore, it can be expected that, for such theories, the junction conditions will have strong influence far beyond the matching hypersurface, and could significantly constrain the fields in the whole spacetime. We will show that this is indeed the case by studying some illustrative examples.

The paper is organized as follows: In Section~\ref{sc:mathprer} we will briefly review distributional calculus on manifolds together with some important formulas for derivatives and distributional curvature tensors. In Section~\ref{sc:idg} we will introduce the action of infinite derivative gravity and present the corresponding field equations. In Section~\ref{sc:matchcond} we will derive the matching conditions on the timelike hypersurface of co-dimension 1 between two regions of the manifold. These conditions are required for the theory to be well-defined in a distributional sense. In Section~\ref{sc:junfieq} we will find expressions for the singular parts and discontinuities of the energy-momentum tensor. We will write the field equations for the quantities on the junction and discuss the proper matching where no matter content is allowed on the hypersurface. In Section \ref{sc:applications} we will study the consequences of our results on several illustrative examples. Particularly, we will examine the implications for the braneworld models and discuss explicit junction conditions for simple star models. The paper will be concluded with a brief summary of our results in Section~\ref{sc:conclusion}.


\section{Mathematical prerequisites}\label{sc:mathprer}

In this section, we give a brief exposition of tensor distributions with application to spacetimes involving surfaces where the geometry is not differentiable. We present only the definitions and important results such as the distributional generalization of the curvature tensors, which are essential for the study of junction conditions. For the derivations and more details, we refer the reader to \cite{Mars:1993mj,Reina:2015gxa}.


\subsection{Tensor distributions}

Consider a 4-dimensional manifold $M$ equipped with a metric $\bs{g}$. Let $\mathcal{D}$ be a space of smooth tensor fields on $M$ with a compact support called the \textit{test tensor fields}. A \textit{tensor distribution} is a linear continuous functional ${\bs{\chi}:\mathcal{D}\to\mathbb{R}}$, which maps test tensor fields ${\bs{Y}\in\mathcal{D}}$ to real numbers,
\begin{equation}
    \bs{\chi}(\bs{Y})\equiv\langle\bs{\chi},\bs{Y}\rangle\;.
\end{equation}
Locally integrable tensor fields $\bs{T}$ define unique tensor distributions $\underline{\bs{T}}$ by means of\footnote{In the index-free notation, we use $\cdot$ to denote contractions of adjacent indices and $\bullet$ to denote multi-contractions in all indices. We also employ the musical isomorphism to raise indices of a 1-forms ${\bs{\alpha}^{\sharp}=\bs{g}^{-1}\cdot\bs{\alpha}}$, and lower indices of vectors ${\bs{v}^{\flat}=\bs{g}\cdot\bs{v}}$.}
\begin{equation}\label{eq:distrassocfields}
    \langle\underline{\bs{T}},\bs{Y}\rangle\equiv\int_M\! \mathfrak{g}^{\frac12} \bs{T}\bullet \bs{Y}\;, \qquad \bs{T}\bullet \bs{Y}\equiv T_{\mu\dots}^{\nu\dots}Y^{\mu\dots}_{\nu\dots}\;,
\end{equation}
where ${\mathfrak{g}^{\frac12}=\sqrt{-g}\,dx^4}$ is the volume element and ${\mu,\nu,\dots=0,1,2,3}$ are the 4-dimensional indices. Note that the action of $\underline{\bs{T}}$ can be extended to non-smooth test fields as well and the action of tensor distribution is considered whenever it can be defined \cite{Mars:1993mj}. Components of a tensor distribution with respect to the vector frame $\bs{e}_{(\mu)}$ (and dual coframe $\bs{e}^{(\nu)}$), $\chi_{\mu\dots}^{\nu\dots}$, are defined by means of
\begin{equation}
     \langle \chi_{\mu\dots}^{\nu\dots} ,Y\rangle \equiv \langle \bs{\chi},Y\bs{e}_{(\mu)}\cdots\bs{e}^{(\nu)}\cdots\rangle\;,
\end{equation}
where $Y$ is a test scalar function. Consequently, we can write ${\langle\bs{\chi},\bs{Y}\rangle=\langle\chi_{\mu\dots}^{\nu\dots},Y^{\mu\dots}_{\nu\dots}\rangle}$\;.

As it is common in the theory of distributions, the operations on distributional objects are defined through the actions on test fields. In particular, the definition of the \textit{covariant derivative of a tensor distribution} is motivated by the integration by parts formula applied to \eqref{eq:distrassocfields},
\begin{equation}\label{eq:covderdistr}
\langle\bs{\nabla}\bs{\chi},\bs{Y}\rangle \equiv -\langle\bs{\chi},\bs{\nabla}\cdot\bs{Y}\rangle\;,\qquad (\bs{\nabla}\cdot\bs{Y})^{\mu\dots}_{\nu\dots}\equiv\nabla_\rho Y^{\rho\mu\dots}_{\nu\dots}\;.
\end{equation}
The tensor \textit{multiplication by a tensor field} is defined by
\begin{equation}
\langle\bs{T}\bs{\chi},\bs{Y}\rangle \equiv \langle\bs{\chi},\bs{T}\bullet\bs{Y}\rangle\;,\qquad (\bs{T}\bullet \bs{Y})^{\alpha\dots}_{\beta\dots}\equiv T_{\mu\dots}^{\nu\dots}Y^{\mu\dots}_{\nu\dots}{}^{\alpha\dots}_{\beta\dots}\;.
\end{equation}
In general, this product is well defined if the tensor field $\bs{T}$ is smooth, however, also  in more general cases \cite{Mars:1993mj}. 

The serious mathematical issue, however, arises with the products of distributions which cannot be associated to tensor fields such as $\delta\delta$ terms \cite{Schwartz}. Note that it is possible to give a mathematical sense to such objects in the theories of nonlinear generalized functions, such as Colombeau algebras \cite{colombeau1990}. Unfortunately, the use of these advanced mathematical tools is quite complicated even within the theory of general relativity \cite{Steinbauer:2006qi,Grosser:1620651}. In this paper, we restrict ourselves to the theory of linear distributions.


\subsection{Matching hypersurface}\label{ssc:matchsurf}
The manifold $M$ possesses two regions ${M^+}$ and ${M^-}$ separated by a 3-dimensional timelike hypersurface $\Sigma$, called the \textit{matching hypersurface}. It is assumed that the metrics in each region, $\bs{g}^{\pm}$, are smooth. To glue these to spacetimes together we have to demand that the 3-dimensional \textit{induced metrics} (a.k.a. the \textit{first fundamental forms}) $\bs{h}^{\pm}$ on $\Sigma$ as viewed from both sides coincide,
\begin{equation}\label{eq:hcoincidence}
h_{ab}^-=h_{ab}^+\equiv h_{ab}\;,
\end{equation}
where ${a,b,\dots=0,1,2}$ are the 3-dimensional indices. This single condition allows us to match the 4-dimensional tangent space on both sides of ${\Sigma}$. By choosing a set of basis vector fields $\bs{e}_{(a)}$ tangent to ${\Sigma}$, we can construct the orthonormal 1-forms $\bs{n}^{\pm}$ whose contraction with $\bs{e}_{(a)}$ vanish
\begin{equation}\label{eq:nn}
    \bs{n}^{\pm}\cdot\bs{e}_{(a)}=0\;,
    \qquad \bs{n}^{\pm}\cdot\bs{n}^{\pm}{}^\sharp=1\;.
\end{equation}
Since such an othonormal 1-form is fixed uniquely up to sign, we can drop the $\pm$ sign and denote the othonormal 1-form pointing towards $M^+$ by $\bs{n}$. The set ${\{\bs{e}_{(a)},\bs{n}^\sharp\}}$ then forms a common frame of both tangent spaces with a dual coframe ${\{\bs{e}^{(a)},\bs{n}\}}$.

We will mostly work with the 4-dimensional version of the induced metric $ h_{\mu\nu}$ that is related to the 3-dimensional $h_{ab}$ one through the relation
\begin{equation}\label{eq:inducedmetric}
    h_{ab}\equiv h_{\mu\nu}e^{\mu}_{(a)} e^{\nu}_{(b)}\;.
\end{equation}
An important consequence of the previous analysis following from \eqref{eq:hcoincidence} is that the complete metric $\bs{g}$ is continuous\footnote{This allows us to freely raise and lower indices everywhere using $\bs{g}$.} across $\Sigma$, but still may have discontinuous finite derivatives. These definitions allows us to decompose an arbitrary tensor into its tangent and normal directions using the formula
\begin{equation}
    g_{\mu\nu}=h_{\mu\nu}+n_{\mu}n_{\nu}\;.
\end{equation}
because $h_{\mu\nu}$ plays a role of the projector to $\Sigma$ thanks to the properties ${h^{\mu}_{\nu}h^{\nu}_{\kappa}=h^{\mu}_{\kappa}}$, ${h^{\mu}_{\nu}n^{\nu}=0}$.


\subsection{Discontinuities and derivatives}
The metric tensor $\bs{g}$ is not differentiable on $\Sigma$, however, it defines a distribution $\underline{\bs{g}}$ which can be differentiated in distributional sense to obtain distributional curvature tensors. In order to proceed in this direction, we need to define the \textit{Heaviside step function} $\theta$ (and the associated distribution $\underline{\theta}$) with the jump on $\Sigma$,
\begin{equation}
    \theta\equiv\begin{cases}
    0\;,\quad & M^-\;,\\
    \tfrac12\;,\quad & \Sigma\;,\\
    1\;,\quad & M^+\;,\\
    \end{cases}
    \qquad
    \langle\underline{{\theta}},Y\rangle =\int_{M^+} \!\! \mathfrak{g}^{\frac12}\,Y\;,
\end{equation}
and \textit{Dirac delta distribution} $\delta$ supported on $\Sigma$,
\begin{equation}
    \langle\delta,Y\rangle \equiv\int_{\Sigma} \mathfrak{h}^{\frac12}\,Y\;,
\end{equation}
where $\mathfrak{h}^{\frac12}$ is the volume element on $\Sigma$. Definition \eqref{eq:covderdistr} together with the divergence theorem implies
\begin{equation}
    \bs{\nabla}\underline{\theta}=\bs{n}\delta\;.
\end{equation}
This relation can be easily generalized to the covariant derivative of a distribution associated to an arbitrary discontinuous tensor field with finite limits and its derivatives from both sides. Consider a tensor field $\bs{T}$ (and the corresponding distribution~${\underline{\bs{T}}}$),
\begin{equation}
    \bs{T}\equiv\bs{T}^+\theta+\bs{T}^-(1-\theta)\;,
    \qquad
    \underline{\bs{T}}=\bs{T}^+\underline{\theta}+\bs{T}^-(\underline{1}-\underline{\theta})\;.
\end{equation}
Its covariant derivative then reads
\begin{equation}\label{eq:dertensordist}
    \bs{\nabla}\underline{\bs{T}}=\bs{\nabla}\bs{T}^+\underline{\theta}+\bs{\nabla}\bs{T}^-(\underline{1}-\underline{\theta})+\bs{n}[\bs{T}]\delta\;,
\end{equation}
where we introduced the square-bracket notation,
\begin{equation}
    [\bs{T}]\equiv\big(\bs{T}^+-\bs{T}^-\big)\big|_{\Sigma}\;,
\end{equation}
to denote the jump of ${\bs{T}}$ across $\Sigma$. We refer to the terms such as $[\bs{T}]\bs{n}\delta$ that prevent the distributions to be associated to a tensor field as the \textit{singular parts of distributions}. Similar to \eqref{eq:dertensordist}, we can also calculate the second derivative, where one additional term appears due to the possible discontinuity of $\bs{\nabla}\bs{T}$,
\begin{equation}
    \bs{\nabla}\bs{\nabla}\underline{\bs{T}}=\bs{\nabla}\bs{\nabla}\bs{T}^+\underline{\theta}+\bs{\nabla}\bs{\nabla}\bs{T}^-(\underline{1}-\underline{\theta})+\bs{\nabla}\big(\bs{n}[\bs{T}]\delta\big)+\bs{n}[\bs{\nabla}\bs{T}]\delta\;.
\end{equation}

Finally, let us mention that we will use the shorthand notation for the value of the $\bs{T}$ on $\Sigma$,
\begin{equation}
    \bs{T}^{\Sigma}\equiv\frac{1}{2}\big(\bs{T}^++\bs{T}^-\big)\big|_{\Sigma}\;.
\end{equation}
The discontinuity of the tensor product of $\bs{A}$ and $\bs{B}$ is then given by
\begin{equation}
    [\bs{A}\bs{B}]=\bs{A}^{\Sigma}[\bs{B}]+[\bs{A}]\bs{B}^{\Sigma}\;.
\end{equation}
The jump of the first derivative of a scalar field $T$ can be written as
\begin{equation}\label{eq:jumpderscal}
    [\nabla_{\mu}T] =n_{\mu}n^{\nu}[\nabla_{\nu}T]+h_{\mu}^{\kappa}\nabla_{\kappa}[T]\;,
\end{equation}
while the second derivative reads (see, e.g., \cite{Reina:2015gxa})
\begin{equation}\label{eq:jumpderderscal0}
    [\nabla_{\mu}\nabla_{\nu}T] = n_{\mu}n^{\kappa}[\nabla_{\kappa}\nabla_{\nu}T] + h_{\mu}^{\kappa}\nabla_{\kappa}[\nabla_{\nu}T]-h_{\mu}^{\rho}[\Gamma^{\kappa}_{\rho\nu}](\nabla_{\kappa}T)^{\Sigma}\;.
\end{equation}
A more convenient form of the last formula can be found by a recursive argument (together with commutation of derivatives of scalars),
\begin{equation}
    n_{\mu}n^{\kappa}[\nabla_{\kappa}\nabla_{\nu}T]+U_{\mu\nu}= n_{\mu}n^{\kappa}\big( n_{\nu}n^{\rho}[\nabla_{\rho}\nabla_{\kappa}T]+U_{\nu\kappa}\big)+U_{\mu\nu}\;,
\end{equation}
where we denoted ${U_{\mu\nu}\equiv h_{\mu}^{\kappa}\nabla_{\kappa}[\nabla_{\nu}T]-h_{\mu}^{\rho}[\Gamma^{\kappa}_{\rho\nu}](\nabla_{\kappa}T)^{\Sigma}}$. After inserting this identity in \eqref{eq:jumpderderscal0}, we arrive at
\begin{equation}\label{eq:jumpderderscal}
    [\nabla_{\mu}\nabla_{\nu}T]=n_{\mu}n_{\nu}n^{\kappa}n^{\rho}[\nabla_{\kappa}\nabla_{\rho}T]+n_{\mu}n^{\kappa}U_{\nu\kappa}+U_{\mu\nu}\;.
\end{equation}


\subsection{Intrinsic and extrinsic curvature}

The fact that we have a metric with discontinuous derivatives across $\Sigma$ means that there can be a discontinuity of the curvature tensors. That is why these tensors need to be described in a distributional sense. Starting with the metric distribution defined by the continuous metric with finite derivatives from both sides,
\begin{equation}
    \underline{g}_{\mu\nu}=g_{\mu\nu}^+\underline{\theta}+g_{\mu\nu}^-(\underline{1}-\underline{\theta})\;,\quad [g_{\mu\nu}]=0\;,
\end{equation}
we can construct distributional Christoffel symbols, 
\begin{equation}
    \underline{\Gamma}^{\kappa}_{\mu\nu}=\Gamma^+{}^{\kappa}_{\mu\nu}\underline{\theta}+\Gamma^-{}^{\kappa}_{\mu\nu}(\underline{1}-\underline{\theta})\;, \quad [\Gamma{}^{\kappa}_{\mu\nu}]\in\mathbb{R}\;,
\end{equation}
using the standard formulas and the rule \eqref{eq:dertensordist}. As a consequence of the arbitrary jump $[\Gamma{}^{\kappa}{}_{\mu\nu}]$, the Riemann tensor distribution,
\begin{equation}\label{eq:riemdistr}
\underline{R}^{\alpha}{}_{\beta\mu\nu}=R^{+\alpha}{}_{\beta\mu\nu}\underline{\theta}+R^{-\alpha}{}_{\beta\mu\nu}(\underline{1}-\underline{\theta})+\mathcal{R}^{\alpha}{}_{\beta\mu\nu}\delta\;,
\end{equation}
contain a term proportional to $\delta$, the singular part of Riemann distribution,
\begin{equation}
\mathcal{R}^{\alpha}{}_{\beta\lambda\mu}=n_{\lambda}\left[\Gamma^{\alpha}_{\beta\mu}\right]-n_{\mu}\left[\Gamma^{\alpha}_{\beta\lambda}\right]\;.
\end{equation}

Unlike the induced metric, the \textit{extrinsic curvatures} of $\Sigma$, denoted by $K_{\mu\nu}^{\pm}$, (a.k.a. the \textit{second fundamental form}), can be different when viewed from both sides. Their 4- and 3-dimensional versions are defined by
\begin{equation}
\label{secondfundamental}
    K_{\mu\nu}^{\pm}\equiv h^{\rho}_{\mu}h^{\sigma}_{\nu}\nabla^{\pm}_{\rho}n_{\sigma}\;,
    \qquad
    K_{ab}^{\pm}\equiv e^{\rho}_{(a)} e^{\mu}_{(b)}  \nabla^{\pm}_{\rho}n_{\mu}\;.
\end{equation}
The Gauss--Codazzi equations also have their two versions. Particularly important are the relations,
\begin{equation}\label{eq:gauscod}
\begin{aligned}
    R^{\pm}-2R^{\pm}_{\mu\nu}n^{\mu}n^{\nu} &=\overline{R}-(K^{\pm})^2+K^{\pm}_{\mu\nu}+K^{\pm}{}^{\mu\nu}\;,
    \\
    n^{\mu}R^{\pm}_{\mu\rho}h^{\rho}_{\nu} &=\overline{\nabla}^{\mu} K_{\mu\nu}^{\pm}-\overline{\nabla}_{\nu} K^{\pm}\;,
\end{aligned}
\end{equation}
where $\overline{\bs{\nabla}}$ is the covariant derivative on $\Sigma$ and $\overline{R}$ is the scalar curvature of $\Sigma$.

There exist an alternative expression for the singular part of Riemann tensor that relates it to the jump of the extrinsic curvature. As shown in \cite{Reina:2015gxa}, one can employ a coordinate transformation and write the jump of the Christoffel symbols as
\begin{equation}\label{eq:jumpchrist}
    [\Gamma^{\alpha}_{\beta\lambda}]=n_{\lambda}[K^{\alpha}_{\beta}]+n_{\beta}[K^{\alpha}_{\lambda}]-n^{\alpha}[K_{\beta\lambda}]\;.
\end{equation}
The we immediately get,
\begin{equation}\label{eq:singriem}
\mathcal{R}_{\alpha\beta\lambda\mu}=-n_{\alpha}\left[K_{\beta\mu}\right]n_{\lambda}+n_{\alpha}\left[K_{\beta\lambda}\right]n_{\mu}-n_{\beta}\left[K_{\alpha\lambda}\right]n_{\mu}+n_{\beta}\left[K_{\alpha\mu}\right]n_{\lambda}\;.
\end{equation}

The Ricci tensor and Ricci scalar distributions can be obtained by contracting \eqref{eq:riemdistr} and \eqref{eq:singriem}\;,
\begin{equation}\label{eq:ricci}
\underline{R}_{\beta \mu}=R_{\beta \mu}^{+} \underline{\theta}+R_{\beta \mu}^{-}(\underline{1}-\underline{\theta})+\mathcal{R}_{\beta \mu} \delta\;,
\qquad
\mathcal{R}_{\beta \mu}\equiv \mathcal{R}^{\rho}{}_{\beta \rho \mu}=-\left[K_{\beta \mu}\right]-\left[K\right] n_{\beta} n_{\mu}\;,
\end{equation}
and
\begin{equation}\label{eq:ricciscalar}
\underline{R}=R^{+} \underline{\theta}+R^{-}(\underline{1}-\underline{\theta})+\mathcal{R} \delta\;,
\qquad \mathcal{R}\equiv \mathcal{R}_{\rho}^{\rho}=-2\left[K\right]\;,
\end{equation}
where we denoted the trace ${K\equiv K^{\mu}_{\mu}}$. Consequently, the Einstein tensor reads
\begin{equation}\label{eq:einstein}
    \underline{G}_{\beta\mu}=G_{\beta\mu}^+\underline{\theta}+G_{\beta\mu}^-(\underline{1}-\underline{\theta})+\mathcal{G}_{\beta\mu}\delta\;,
    \qquad \mathcal{G}_{\beta\mu}\equiv -[K_{\beta\mu}]+[K]h_{\beta\mu}\;.
\end{equation}
In the next section we shall use these expressions to see how the action and the field equations behave in a divided spacetime, in order to obtain the matching conditions in the boundary. Using the Gauss--Codazzi equations \eqref{eq:gauscod}, it is possible to derive geometric relations,
\begin{equation}\label{eq:israeleinstein}
    K^{\Sigma}_{\rho\sigma}\mathcal{G}^{\rho\sigma}=n^{\beta}n^{\mu}[G_{\beta\mu}]=n^{\beta}n^{\mu}[R_{\beta\mu}]-[R]\;, \qquad \overline{\nabla}^{\beta}\mathcal{G}_{\beta\mu}=-n^{\rho}h^{\sigma}_{\mu}[G_{\rho\sigma}]=-n^{\rho}h^{\sigma}_{\mu}[R_{\rho\sigma}]\;.
\end{equation}
In GR, when Einstein's equations are imposed, these equations are known as \textit{Israel's junction equations} \cite{Israel:1966rt}. It is also possible to derive distributional Bianchi identities \cite{Mars:1993mj},
\begin{equation}
    \nabla_{\rho}\underline{R}^{\alpha}{}_{\beta\mu\nu}+\nabla_{\mu}\underline{R}^{\alpha}{}_{\beta\nu\rho}+\nabla_{\nu}\underline{R}^{\alpha}{}_{\beta\rho\nu}=0\;, \qquad \nabla^{\beta}\underline{G}_{\beta\mu}=0\;.
\end{equation}


\section{Infinite derivative gravity}\label{sc:idg}

Consider the following non-local gravity theory, which we refer to as the \textit{infinite derivative gravity} (IDG) if not stated otherwise\footnote{We consider ${R+RF\left(\Box\right)R}$ theory as a simplification of the complete infinite derivative theory, which also includes quadratic terms with Ricci and Riemann tensors \cite{Biswas:2011ar}. This simplified action was studied in the cosmological context in \cite{Biswas:2005qr}.},
\begin{equation}\label{eq:action}
S=\frac{1}{2}\int_M \mathfrak{g}^{\frac12}\big(R+RF\left(\Box\right)R\big)+S_{\textrm{m}}\;,
\end{equation}
where the analytic operator $F\left(\Box\right)$, often referred to as the \textit{form-factor}, is defined as
\begin{equation}
F\left(\Box\right)=\sum_{n=0}^\infty f_{n}\Box^{n}.
\end{equation}
We hide the parameter $\ell$ describing the \textit{length scale of non-locality} in the coefficients $f_{n}$,
\begin{equation}
    f_{n} =\ell^{2n+2}\hat{f}_{n}\;.
\end{equation}
Furthermore, we assume that $F(\Box)$ is non-polynomial which is reflected in the presence of the infinite number of derivatives, and thus the non-local theory. In other words, for every ${k\geq 0}$ there exists ${n>k}$ such that ${f_n\neq 0}$.

Performing variations with respect to the metric we can find the field equations (see \cite{Biswas:2005qr,Biswas:2013cha}),
\begin{equation}\label{eq:fieldeq}
G_{\mu\nu}+2R_{\mu\nu}F\left(\Box\right)R-2\left(\nabla_{\mu}\nabla_{\nu}{-}g_{\mu\nu}\Box\right)F\left(\Box\right)R-\frac12 g_{\mu\nu}RF\left(\Box\right)R-R\overleftrightarrow{\Omega}_{\mu\nu}R+\frac12 g_{\mu\nu}\big(R\overleftrightarrow{\Omega}_{\sigma}^{\sigma}R+R\overleftrightarrow{\Theta}\Box R\big)=T_{\mu\nu},
\end{equation}
where $\overleftrightarrow{\bs{\Omega}}$ and $\overleftrightarrow{\Theta}$ are bi-linear operators with the left/right actions
defined by
\begin{equation}\label{eq:omega}
\begin{aligned}
    \phi\overleftrightarrow{\Omega}_{\mu\nu} \psi&\equiv\phi\overrightarrow{\nabla}_{\nu}\frac{F(\overleftarrow{\Box})-F(\overrightarrow{\Box})}{\overleftarrow{\Box}-\overrightarrow{\Box}}\overleftarrow{\nabla}_{\mu}\psi & &\equiv \sum_{n=1}^{\infty}f_{n}\sum_{k=0}^{n-1}\nabla_{\mu}\Box^k\phi\nabla_{\nu}\Box^{n-k-1}\psi & &=\sum_{k=0}^{\infty}\nabla_{\mu}\Box^k\phi\sum_{l=0}^{\infty}f_{k+l+1}\nabla_{\nu}\Box^{l}\psi\;,
    \\
    \phi\overleftrightarrow{\Theta}\psi &\equiv \phi\frac{F(\overleftarrow{\Box})-F(\overrightarrow{\Box})}{\overleftarrow{\Box}-\overrightarrow{\Box}}\psi & &\equiv\sum_{n=1}^{\infty}f_{n}\sum_{k=0}^{n-1}\Box^k\phi\Box^{n-k-1}\psi & &=\sum_{k=0}^{\infty}\Box^k\phi\sum_{l=0}^{\infty}f_{k+l+1}\Box^{l}\psi\;,
\end{aligned}
\end{equation}
and $T_{\mu\nu}$ is the usual energy-momentum tensor.


\section{Matching conditions}\label{sc:matchcond}

The moment one considers quadratic curvature terms in the action, problematic terms such as $\delta\delta$ may appear. These kinds of expressions are \textit{ill defined} in the theory of distributions. Instead of dealing with the complicated non-linear generalized functions, we adopt the pragmatic approach of \cite{Mars:1993mj,Deruelle:2007pt,Reina:2015gxa}. It is shown that the standard theory of distributions can be used if the problematic products of distributions are avoided by imposing certain minimal conditions on spacetimes in consideration. In this section we find such conditions.


\subsection{Simplifying assumptions}\label{ssc:simplassump}
Following \cite{Mars:1993mj,Senovilla:2013vra,Reina:2015gxa}, we need to find restrictions on various quantities in $\Sigma$ so that the action and the field equations of the theory are mathematically sensible. For example, in the local limiting case ${F\left(\Box\right)=1}$ one is forced to impose that the jump of the trace of the extrinsic curvature is zero, ${[K]=0}$, see \cite{Senovilla:2013vra}. In the infinite derivative gravity it is not straightforward to generalize this argument. This is due to the fact that the differential operator $F\left(\Box\right)$ can in principle regularize $\delta$ distribution or de-regularize a smooth function. Such situations are difficult to control.

We place a simplifying assumption on the class of form-factors $F(\Box)$ we want to study in a given background.
Roughly speaking, we assume that the action of the operators such as\footnote{A typical form factor for a ghost-free IDG around the Minkowski space time is given by 
$F(\Box)=\frac{e^{-\ell^2\Box}-1}{\Box}$ ~\cite{Biswas:2005qr,Biswas:2013cha}, which does not produce a regular function when it acts on $\delta$ distribution localized at timelike hypersuface in flat Minkowski spacetime. }
\begin{equation}\label{eq:assumption}
\sum_{n=0}^\infty f_{n+k}\Box^{n}\;,
\qquad
\sum_{n=0}^\infty f_{n+k}\nabla_{\mu}\Box^{n}\;,
\qquad
k\geq0\;,
\end{equation}
on tensor distributions with non-zero singular parts should not give rise to tensor distributions without singular parts. On top of that, we assume that the result of such actions on perfectly smooth tensor fields should not produce tensor distributions with non-zero singular parts. This means that not only $F(\Box)$ but also the one-sided action of $\overleftrightarrow{\bs{\Omega}}$ and $\overleftrightarrow{\Theta}$ do not regularize any singular parts of distributions or de-regularize any smooth tensor fields that appear in our discussion. These rather imprecise demands should become clearer from our construction of the junction conditions in Section~\ref{ssc:illfe}. They are necessary to assume since we want to be able to impose the conditions by avoiding ill-defined expressions in the infinite sums separately for each term only without worrying about convergence of smooth functions to distributions and vice versa.


\subsection{Derivatives of $\underline{R}$}
The action as well as the field equations contain various terms with covariant derivatives of the Ricci scalar. After a repeated use of distributional formula \eqref{eq:dertensordist}, we arrive at
\begin{equation}
\begin{aligned}
    \nabla_{\mu_k}\cdots\nabla_{\mu_1} \underline{R} &= \nabla_{\mu_k}\cdots\nabla_{\mu_1}R^+\underline{\theta}+\nabla_{\mu_k}\cdots\nabla_{\mu_1}R^-(\underline{1}-\underline{\theta})
    \\
    &\feq+\sum_{l=2}^{k+1}\nabla_{\mu_k}\cdots\nabla_{\mu_l}\big(n_{\mu_{l-1}}[\nabla_{\mu_{l-2}}\cdots\nabla_{\mu_1}R]\delta\big)
    +\nabla_{\mu_k}\cdots\nabla_{\mu_1} \big(\mathcal{R}\delta\big)\;.
\end{aligned}
\end{equation}
Setting (${k=2j}$) and contracting pairs of adjacent indices we find the formula for an arbitrary power of $\Box$,
\begin{equation}\label{eq:boxj}
    \Box^j\underline{R}=\Box^j R^+\underline{\theta} +\Box^j R^-(\underline{1}-\underline{\theta}) +\sum_{l=0}^{j-1}\Box^{l}\big(n^{\nu}[\nabla_{\nu}\Box^{j-1-l}R]\delta\big)+\sum_{l=0}^{j-1}\Box^l\nabla_{\nu}\big(n^{\nu}[\Box^{j-1-l}R]\delta\big)+\Box^j(\mathcal{R}\delta)\;.
\end{equation}
A different contraction (with ${k=2j+1}$) or a covariant differentiation of the last equation lead to
\begin{equation}\label{eq:nabboxj}
\begin{aligned}
    \nabla_{\mu}\Box^j\underline{R} &=\nabla_{\mu}\Box^j R^+\underline{\theta} +\nabla_{\mu}\Box^j R^-(\underline{1}-\underline{\theta})+n_\mu[\Box^j R]\delta+\sum_{l=0}^{j-1}\nabla_{\mu}\Box^l\big(n^{\nu}[\nabla_{\nu}\Box^{j-1-l}R]\delta\big)
    \\
    &\feq+\sum_{l=0}^{j-1}\nabla_{\mu}\Box^l\nabla_{\nu}\big(n^{\nu}[\Box^{j-1-l}R]\delta\big)+\nabla_{\mu}\Box^j(\mathcal{R}\delta)\;.
\end{aligned}
\end{equation}


\subsection{Well-defined theory in a distributional sense}\label{ssc:illfe}
Field equations \eqref{eq:fieldeq} contain several terms that are quadratic in curvature:
\begin{equation}\label{eq:terms}
    R_{\mu\nu} F(\Box)R\;,
    \qquad
    R F(\Box)R\;,
    \qquad
    R\overleftrightarrow{\Omega}_{\mu\nu}R\;,
    \qquad
    R\overleftrightarrow{\Omega}_{\sigma}^{\sigma} R\;,
    \qquad
    R\overleftrightarrow{\Theta}\Box R\;.
\end{equation}
Here, we ignored the numerical factors and the metric  $g_{\mu\nu}$, which is continuous and does not affect our discussion. Due to our simplifying assumptions in Section~\ref{ssc:simplassump}, the singular parts of the Ricci scalar distribution cannot be regularized due to the presence of infinite sums \eqref{eq:assumption}. Also, the action of operators on smooth $R$ cannot suddenly produce singular terms. This means that the conditions can be imposed on each summand separately. Furthermore, we have to ignore all dubious cancellations between ill-defined terms, such as $${a_1\delta\delta+a_2\delta\delta=(a_1+a_2)\delta\delta},$$ and simply demand every coefficient to be zero, ${a_1=a_2=0}$, so that no ill-defined terms ever appear (see \cite{Reina:2015gxa}). Such cancellation, if it really occurs, could only be addressed in theory of non-linear generalized functions.

As the conditions arising from the presence of $\overleftrightarrow{\bs{\Omega}}$ and $\overleftrightarrow{\Theta}$~terms are very restrictive, it is best to start with terms involving these operator and then return to the rest. The expressions proportional to $f_n$ of the last three terms in \eqref{eq:terms} read, respectively,
\begin{equation}\label{eq:threeterms}
    \sum_{k=0}^{n-1}\nabla_{\mu}\Box^k R\nabla_{\nu}\Box^{n-k-1} R\;,
    \qquad
    \sum_{k=0}^{n-1}\nabla_{\mu}\Box^k R\nabla^{\mu}\Box^{n-k-1} R\;,
    \qquad
    \sum_{k=0}^{n-1}\Box^k R \Box^{n-k} R\;.
\end{equation}
These terms can make sense in the theory of distributions if the singular parts of ${\nabla_{\mu}\Box^k \underline{R}}$ for ${0\leq k\leq\lfloor\frac{n-1}{2}\rfloor}$ and singular parts of ${\Box^k \underline{R}}$ for ${0\leq k\leq\lceil\frac{n-1}{2}\rceil}$ vanish. After realizing that a non-polynomial $F(\Box)$ guarantees the existence of ${f_n\neq 0}$ with $n$ larger than any arbitrary number and employing \eqref{eq:ricciscalar}, \eqref{eq:boxj}, and \eqref{eq:nabboxj}, these requirements can be achieved by imposing
\begin{equation}\label{eq:condit}
    [K]=0\;,
    \qquad
    [\Box^k R]=0\;,
    \qquad
    n^{\mu}[\nabla_{\mu}\Box^k R]=0\;, \qquad k\geq 0\;.
\end{equation}
In fact, the last condition can be strengthen with the help of \eqref{eq:jumpderscal},
\begin{equation}\label{eq:conditions}
    \boxed{[K]=0\;,
    \qquad
    [\Box^k R]=0\;,
    \qquad
    [\nabla_{\mu}\Box^k R]=0\;, \qquad k\geq 0\;.}
\end{equation}
Taking into consideration the first two conditions of \eqref{eq:conditions}, it is now easy to see that $F(\Box)\underline{R}$ does not contain the singular part. Therefore, the first two terms of \eqref{eq:terms} are well defined as well. The equations \eqref{eq:conditions} are the \textit{junction conditions of IDG}. They can be compared with local ${R+R^2}$ theory, $F(\Box)=1$, which is well defined whenever the trace of extrinsic curvature has to be zero, ${[K]=0}$, but $R$ can be discontinuous, ${[R]\neq 0}$, and its higher derivatives contain singular parts.


Since the action \eqref{eq:action} contains just the quadratic term,
\begin{equation}
    R F(\Box)R\;,
\end{equation}
which was studied above, the conditions \eqref{eq:conditions} are already sufficient to make the action mathematically sensible. Let us note that also the intermediary expressions that are produced in the integration by parts, ${R(\Box^n R) \to (\Box^n R)R}$,
\begin{equation}
    \nabla_{\mu}\Box^k R\nabla^{\mu}\Box^{n-k-1}R\;, 
    \qquad
    \Box^k R\Box^{n-k}R\;,
    \qquad 
    0\leq k\leq n-1\;,
\end{equation}
are well defined, since these are exactly the terms that were freed from ill-defined expressions in \eqref{eq:threeterms}.


\section{Junction field equations}\label{sc:junfieq}
Having obtained the junctions conditions \eqref{eq:conditions}, we can now ask what matter content is allowed on the hypersurface~$\Sigma$. We identify the parts of the energy momentum tensor that are associated to $\Sigma$, derive the respective junction field equations, and discuss the conditions where there is no matter concentrated on $\Sigma$.


\subsection{Singular parts of ${T_{\mu\nu}}$}
Following \cite{Senovilla:2014kua,Reina:2015gxa}, let us consider an energy-momentum tensor distribution with a singular part proportional to~$\delta$,
\begin{equation}
    \underline{T}_{\mu\nu}=T_{\mu\nu}^{+} \underline{\theta}+T_{\mu\nu}^{-}(\underline{1}-\underline{\theta})+\mathcal{T}_{\mu\nu}\delta\;,
\end{equation}
whose coefficient can be decomposed into the tangent, tangent-normal, and normal parts,
\begin{equation}\label{eq:tdecomp}
\begin{gathered}
    \mathcal{T}_{\mu\nu}\equiv\tau_{\mu\nu}+\rho_{\mu}n_{\nu}+\rho_{\nu}n_{\mu}+\sigma n_{\mu}n_{\nu}\;,
    \\
    \tau_{\mu\nu}\equiv h_{\mu}^{\rho}h_{\nu}^{\sigma}\mathcal{T}_{\rho\sigma}\;,
    \quad
    \rho_{\mu}\equiv h_{\mu}^{\rho} n^{\sigma}\mathcal{T}_{\rho\sigma}\;,
    \quad
    \sigma\equiv n^{\rho} n^{\sigma}\mathcal{T}_{\rho\sigma}\;.
\end{gathered}
\end{equation}
Here, $\tau_{\mu\nu}$ is the \textit{energy-momentum tensor on $\Sigma$}, $\rho_{\mu}$ is the \textit{external flux momentum}, and $\sigma$ is the \textit{external tension}. Assuming that the conditions \eqref{eq:conditions} are satisfied, we can now return to the field equations \eqref{eq:fieldeq}. Thanks to \eqref{eq:ricci} and \eqref{eq:einstein}, only the first two terms of \eqref{eq:fieldeq} have non-trivial singular part,
\begin{equation}
\begin{aligned}
    \underline{G}_{\mu\nu} &=G_{\mu\nu}^+\underline{\theta}+G_{\mu\nu}^-(\underline{1}-\underline{\theta})+\mathcal{G}_{\mu\nu}\delta\;,
    \\
    \underline{R}_{\mu\nu} F(\Box)\underline{R} &=R_{\mu\nu}^+F(\Box)R^+\underline{\theta}+R_{\mu\nu}^-F(\Box)R^-(\underline{1}-\underline{\theta})+\mathcal{R}_{\mu\nu}(F(\Box)R)^{\Sigma}\delta\;.
\end{aligned}
\end{equation}
Indeed, the singular part of ${\nabla_{\mu}\nabla_{\nu}F(\Box)\underline{R}}$ is zero since it equals ${\nabla_{\mu}\big(n_{\nu}[F(\Box)R]\delta\big)+n_{\mu}[\nabla_{\nu}F(\Box)R]\delta=0}$. Terms with  $\overleftrightarrow{\bs{\Omega}}$ and $\overleftrightarrow{\Theta}$ vanish because all expressions in the products are free from singular parts. Thus, we arrive at
\begin{equation}
    \mathcal{T}_{\mu\nu} =\mathcal{G}_{\mu\nu}+2\mathcal{R}_{\mu\nu}(F(\Box)R)^{\Sigma}\;.
\end{equation}
Employing \eqref{eq:ricci}, \eqref{eq:einstein}, and \eqref{eq:tdecomp}, we find that the theory admits the non-zero energy-momentum tensor on $\Sigma$ with vanishing external flux momentum and external tension,
\begin{equation}\label{eq:tautautau}
    \boxed{\tau_{\mu\nu} =-\big(1+2(F(\Box)R)^{\Sigma}\big)\left[K_{\mu \nu}\right]\;,
    \quad
    \rho_{\mu} =0\;,
    \quad
    \sigma =0\;.}
\end{equation}
This differs from ${R+R^2}$ theory by vanishing external energy flux and external tension and from GR (for which ${\rho_{\mu}=\sigma=0}$) by a different form of the energy-momentum tensor on $\Sigma$, which is ${\tau_{\mu\nu}=-[K_{\mu\nu}]+h_{\mu\nu}[K]}$. An interesting property of ${\tau_{\mu\nu}}$ in IDG is that it is necessarily trace-less, ${\tau\equiv\tau_{\mu}^{\mu}=0}$.


\subsection{Jumps of ${T_{\mu\nu}}$}
Let us find the equations for the jump of $T_{\mu\nu}$. Subtracting the field equations on both sides of $\Sigma$,
\begin{equation}
\begin{aligned}
    T_{\mu\nu}^{\pm} &= G_{\mu\nu}^{\pm}+2R_{\mu\nu}^{\pm}F\left(\Box\right)R^{\pm}-2\left(\nabla_{\mu}\nabla_{\nu}-g_{\mu\nu}^{\pm}\Box\right)F\left(\Box\right)R^{\pm}-\frac12 g_{\mu\nu}^{\pm}R^{\pm}F\left(\Box\right)R^{\pm}
    \\
    &\feq-R^{\pm}\overleftrightarrow{\Omega}_{\mu\nu}R^{\pm}+\frac12 g_{\mu\nu}^{\pm}\big(R^{\pm}\overleftrightarrow{\Omega}_{\sigma}^{\sigma}R^{\pm}+R^{\pm}\overleftrightarrow{\Theta}\Box R^{\pm}\big)\;,
\end{aligned}
\end{equation}
we obtain
\begin{equation}\label{eq:jumpTmunu}
    [T_{\mu\nu}] =[G_{\mu\nu}]+2[R_{\mu\nu}](F(\Box)R)^{\Sigma}-2[\nabla_{\mu}\nabla_{\nu}F(\Box)R]\;.
\end{equation}
The jump of the 2nd covariant derivative can be calculated using \eqref{eq:jumpderderscal}, where $U_{\mu\nu}$ reduces to ${-h_{\mu}^{\rho}[\Gamma^{\kappa}_{\rho\nu}](\nabla_{\kappa}F(\Box)R)^{\Sigma}}$, and the formula for the jump of Christoffel symbols \eqref{eq:jumpchrist},
\begin{equation}\label{eq:nnFR}
\begin{aligned}
    [\nabla_{\mu}\nabla_{\nu}F(\Box)R]  &=n_{\mu}n_{\nu}n^{\kappa}n^{\rho}[\nabla_{\kappa}\nabla_{\rho}F(\Box)R]-\big(n_{\mu}[K_{\nu}^{\lambda}]+n_{\nu}[K_{\mu}^{\lambda}]-n^{\lambda}[K_{\mu\nu}]\big)(\nabla_{\lambda}F(\Box)R)^{\Sigma}
    \\
    &=-\big(n_{\mu}[K_{\nu}^{\lambda}]+n_{\nu}[K_{\mu}^{\lambda}]-n^{\lambda}[K_{\mu\nu}]\big)(\nabla_{\lambda}F(\Box)R)^{\Sigma}\;.
\end{aligned}
\end{equation}
where the second equality comes from the trace of the first line. The energy-momentum quantities \eqref{eq:tautautau} arise from the normal and tangent-normal projections of the right-hand side of \eqref{eq:jumpTmunu}. Employing relation \eqref{eq:nnFR} together with~\eqref{eq:israeleinstein},
\begin{equation}
\begin{aligned}
    n^{\mu}n^{\nu}[G_{\mu\nu}] &=n^{\mu}n^{\nu}[R_{\mu\nu}]=-K_{\mu\nu}^{\Sigma}[K^{\mu\nu}]\;,
    \\
    n^{\mu}h^{\nu}_{\kappa}[G_{\mu\nu}] &=n^{\mu}h^{\nu}_{\kappa}[R_{\mu\nu}]=\overline{\nabla}^{\mu}[K_{\mu\kappa}]\;,
\end{aligned}
\end{equation}
we arrive at the \textit{junction field equation of IDG},
\begin{equation}\label{eq:junceq}
\begin{aligned}
    n^{\mu}n^{\nu}[T_{\mu\nu}] &= -\big(1+2(F(\Box)R)^{\Sigma}\big)K^{\Sigma}_{\mu\nu}[K^{\mu\nu}]\;,
    \\
    n^{\mu}h^{\nu}_{\kappa}[T_{\mu\nu}] &=\big(1+2(F(\Box)R)^{\Sigma}\big)\overline{\nabla}^{\mu}[K_{\mu\kappa}]+2[K_{\kappa}^{\mu}](\nabla_{\mu}F(\Box)R)^{\Sigma}\;.
\end{aligned}
\end{equation}
Since the divergence  of \eqref{eq:tautautau} reads
\begin{equation}
    \overline{\nabla}^{\mu}\tau_{\mu\kappa}=-\big(1+2(F(\Box)R)^{\Sigma}\big)\overline{\nabla}^{\mu}K_{\mu\kappa}-2[K_{\mu\kappa}]\overline{\nabla}^{\mu}\big((F(\Box)R)^{\Sigma}\big)\;,
\end{equation}
these junction field equations can be rewritten purely in terms of the energy-momentum tensor on $\Sigma$,
\begin{equation}\label{eq:juncfieeq}
\begin{aligned}
    n^{\mu}n^{\nu}[T_{\mu\nu}] &= K^{\Sigma}_{\mu\nu}\tau^{\mu\nu}\;,
    \\
    n^{\mu}h^{\nu}_{\kappa}[T_{\mu\nu}] &=-\overline{\nabla}^{\mu}\tau_{\mu\kappa}\;.
\end{aligned}
\end{equation}
In this form, the junction field equations resemble their GR counterparts. The difference is in the expression for the energy-momentum quantities on $\Sigma$ which is given by \eqref{eq:tautautau} for IDG.


\subsection{Proper matching conditions}

Let us discuss a physically important scenario in which no matter content is allowed on the matching hypersurface $\Sigma$. This situation corresponds to the vanishing singular part of the energy-momentum tensor, ${\mathcal{T}_{\mu\nu}=0}$. The resulting conditions are often called the \textit{proper junction conditions}, see \cite{Reina:2015gxa}. The most studied examples of this kind involve the matching of the interior and exterior geometries for various static or dynamical star with models. In our specific theory, the requirement that ${\mathcal{T}_{\mu\nu}=0}$ implies that either the jump of the extrinsic curvature vanishes (see \eqref{eq:tautautau}),
\begin{equation}\label{eq:K0}
\left[K_{\mu \nu}\right]=0\;,
\end{equation}
or the $F(\Box)$-dependent expression
\begin{equation}\label{eq:oneplus2f}
1+2(F(\Box)R)^{\Sigma}=0\;,
\end{equation}
is zero.

The first condition leads to the Ricci and Riemmann tensors that are free of singular parts. The corresponding proper matching conditions then read
\begin{equation}\label{eq:properconditions}
    \boxed{\left[K_{\mu \nu}\right]=0\;,
    \qquad
    [\Box^k R]=0\;,
    \qquad
    [\nabla_{\mu}\Box^k R]=0\;, \qquad k\geq 0\;.}
\end{equation}
It is worth noting that the extra condition \eqref{eq:K0} that arises from the proper matching is the same as in GR. This is because the term this theory introduces to the singular part of the energy-momentum tensor is directly proportional to the step of the extrinsic curvature. This is not true, for example, in modifications of GR that also involve quadratic terms of the Riemann tensor \cite{Reina:2015gxa}, where the proper matching leads to vanishing jumps (and first derivatives) of the Riemann tensor.

The second option \eqref{eq:oneplus2f} can be rewritten as 
\begin{equation}
    F(\Box)R^{\pm}\big|_{\Sigma}=-\frac12\;,
\end{equation}
where we used ${[\Box^k R]=0}$. Unfortunately, this equation requires evaluation of the action of a $F(\Box)$, which is highly non-trivial in a generic spacetime. However, it might be worth studying in cases where $R$ can be decomposed into a sum of eigenfunctions of $\Box$.


\section{Applications}\label{sc:applications}

\subsection{Implications for braneworld models}
Braneworld models of gravity are constructed to solve the hierarchy problem and the cosmological constant problem \cite{Maartens:2010ar,Flanagan:2000nx}. The most common example is the Randall--Sundrum braneworld \cite{Randall:1999ee,Randall:1999vf}. Such a model consists of a 5-dimensional anti-de Sitter spacetime (called the \textit{bulk}) with $\mathbb{Z}_2$-symmetry  with respect to a flat 4-dimensional hypersurface (referred to as the \textit{brane}) representing our 4-dimensional world. Following the covariant Shiromizu--Maeda--Sasaki approach \cite{Shiromizu:1999wj} (see also \cite{Maartens:2010ar} for a review), one can generalize this study to any 5-dimensional bulk geometry with $\mathbb{Z}_2$ symmetry with respect to an arbitrary 4-dimensional brane, and to arbitrary gravity theories in the bulk. The bulk metric is assumed to be a solution of the considered theory with a cosmological constant, and the brane field equations arise from the junction conditions of the theory. This is where our results come into play, if the 5-dimensional bulk theory is taken to be IDG.

The equation \eqref{eq:tautautau} has a direct physical consequence on possible cosmological braneworld models. In order to explore the properties of the brane in IDG, we write its energy-momentum tensor as\footnote{In this section only, we use ${\mu,\nu,\dots=0,1,2,3,4}$ for the bulk indices and ${a,b,\dots=0,1,2,3}$ for the brane indices.}
\begin{equation}\label{eq:EMbrane}
    \tau_{ab}=\tau_{ab}^{\textrm{m}}-\lambda h_{ab}\;,
\end{equation}
where $\tau_{ab}^{\textrm{m}}$ represents the energy-momentum tensor of the matter fields confined to the brane, and $\lambda$ is the \textit{internal tension of the brane}\footnote{Do not confuse with the external tension $\sigma$.}, which accounts for the accelerated cosmological-constant-like expansion on the brane. In the original Randall--Sundrum model in GR, it is related to the cosmological constant in the bulk.

Taking the trace of \eqref{eq:EMbrane}, we obtain
\begin{equation}
    h^{ab}\tau_{ab}=\tau^{\textrm{m}}-4\lambda\;,
\end{equation}
where ${\tau^{\textrm{m}}\equiv h^{ab}\tau_{ab}^{\textrm{m}}}$. Clearly, the left-hand side is actually the trace of the total energy-momentum tensor on the brane~$\tau_{\mu\nu}$ (cf. \eqref{eq:tdecomp}),
\begin{equation}
    \tau=\tau^{\textrm{m}}-4\lambda\;,
\end{equation}
because ${\tau=g^{\mu\nu}\tau_{\mu\nu}=h^{\mu\nu}\tau_{\mu\nu}=\tau_{\mu\nu}h^{ab}e^{\mu}_{(a)}e^{\mu}_{(b)}=h^{ab}\tau_{ab}}$. Finally, resorting to IDG in the bulk, we know that the trace of the total energy-momentum is zero, ${\tau=0}$, which allows to write the following condition on the internal tension 
\begin{equation}
    \lambda=\frac{\tau^{\textrm{m}}}{4}\;.
\end{equation}

\begin{figure}
    \centering
    \includegraphics[scale=0.45]{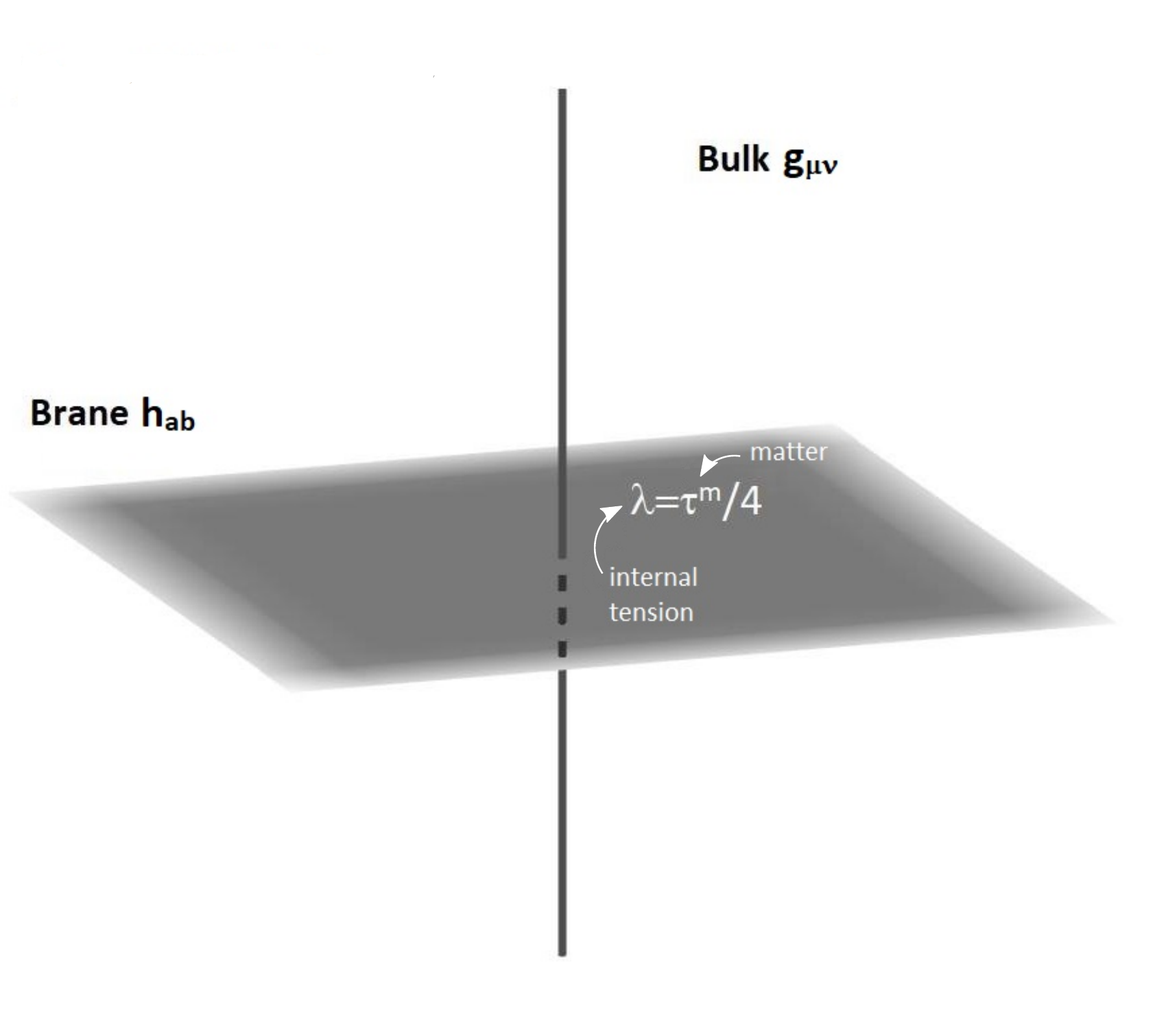}
    \caption{Consequences of the junction conditions for the braneworld models with the non-local ${R{+}RF\left(\Box\right)R}$ theory in the 5-dimensional bulk. The internal tension $\lambda$ of the 4-dimensional brane is given purely by the trace of the energy-momentum tensor of the matter confined to the brane ${\tau^{\textrm{m}}\equiv h^{ab}\tau_{ab}^{\textrm{m}}}$.}
    \label{fig:braneworld}
\end{figure}

Therefore, the internal tension is given purely by the trace of the stress-energy tensor of the matter confined to the brane, as it ís depicted in Figure~\ref{fig:braneworld}. The fact that the matter fields in the brane fix the value of the tension has huge consequences in the possible braneworld models with this non-local theory in the bulk. For example, the original Randall--Sundrum model, which assumes no matter content, cannot describe the accelerated expansion on the brane. On the other hand, generalisations of such a model could resolve the fine-tuning problem of the cosmological constant in the bulk \cite{Forste:2000ge}, since now the tension is uniquely related to the matter content on the brane.

\subsection{Junction conditions in star models}

\begin{figure}
    \centering
    \includegraphics[scale=0.3]{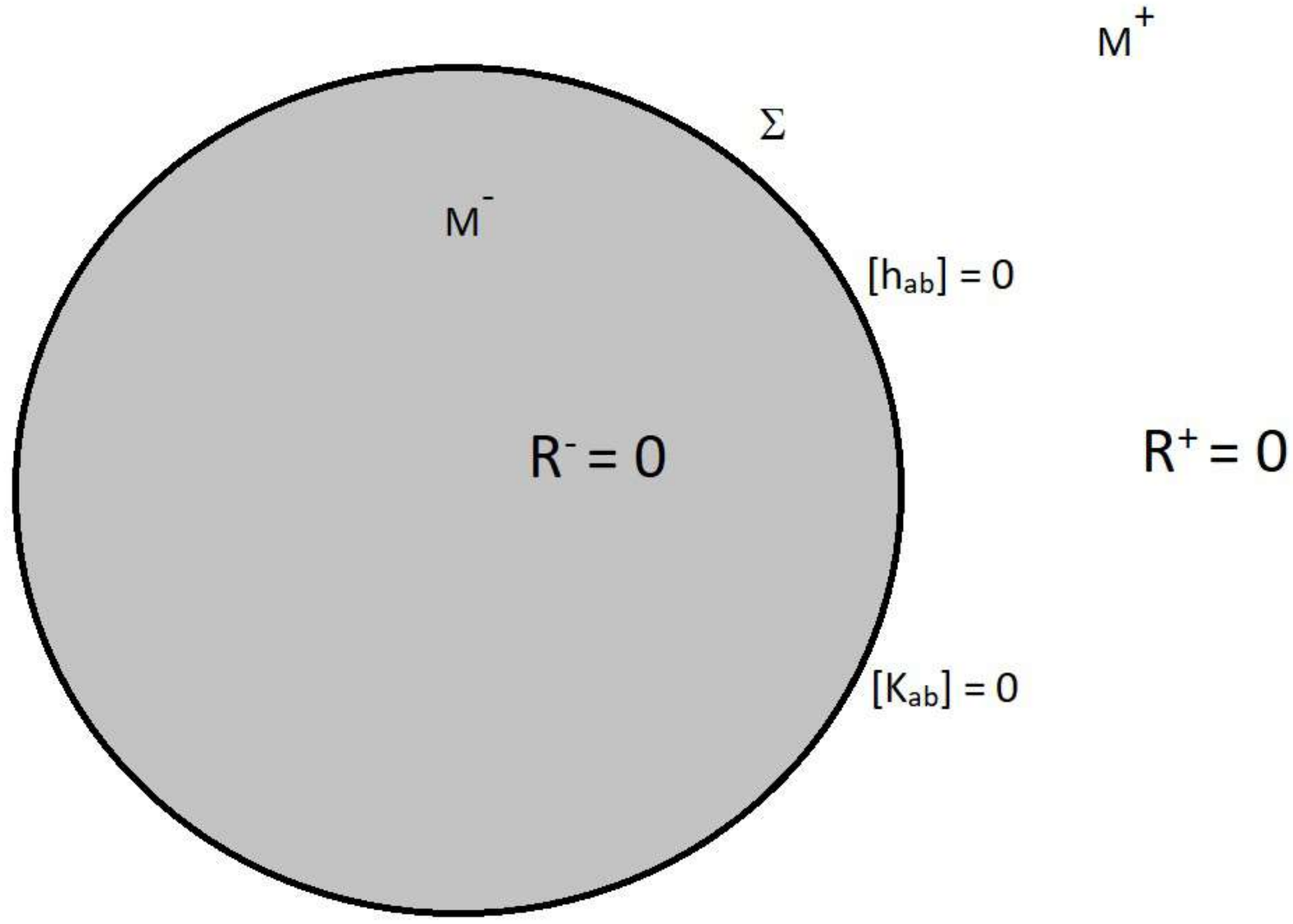}
    \caption{Consequences of the junction conditions in the non-local ${R{+}RF\left(\Box\right)R}$ theory for star models with vanishing Ricci scalar in the exterior, ${R^{+}=0}$. The infinite number of conditions implies that the Ricci scalar is zero also in the interior,~${R^{-}=0}$.}
    \label{fig:starmodels}
\end{figure}

The fact that the proper junction conditions \eqref{eq:properconditions} impose so many constraints strongly affects the possible metrics that could potentially model physical situations by matching the geometries on the junction between $M^+$ and $M^-$. In order to illustrate this, we study two paradigmatic examples. First, we consider a generic static and spherically symmetric star whose exterior region has vanishing Ricci scalar. Then, we discuss a collapsing star with an interior region modelled by the Friedmann--Lemaitre--Robertson--Walker metric (FLRW) and the spherically symmetric static exterior with vanishing Ricci scalar. It turns out that in both cases the junction conditions hold only if the Ricci scalar is zero everywhere, as is illustrated schematically in Figure~\ref{fig:starmodels}. 
\begin{itemize}
\item This implies that most of the GR solutions that describe the interior of static stars or gravitational collapse are not compatible with a vanishing Ricci scalar exterior in this non-local theory.
\end{itemize}
Finally, let us stress that the main purpose of this section is to demonstrate the power of the junction conditions only, not to find new non-local solutions, since it would require solving highly complicated field equations of IDG.


\subsubsection{Static star}

Let us consider a model of a static spherically symmetric star with the exterior region~$M^+$ and the interior region~${M^-}$ separated by a timelike hypersurface $\Sigma$. The geometry on both sides is described by the metrics (see, e.g., \cite{hawking_ellis_1973})
\begin{equation}
\bs{g}^{\pm}=-p^{\pm}(r)\bs{\dd} t^{2}+q^{\pm}(r)\bs{\dd} r^{2}+r^{2}\left(\bs{\dd} \vartheta^{2}+\sin^{2}\vartheta\,\bs{\dd} \varphi^{2}\right)\;,
\end{equation}
where ${p^{\pm}(r)>0}$ and ${q^{\pm}(r)>0}$ are four single-variable functions. The junction hypersurface $\Sigma$ is located at ${r=r_0}$. It is clear that one indispensable condition from the continuity of the metric is 
\begin{equation}
p^-(r_0)=p^+(r_0)\;,
\quad
q^-(r_0)=q^+(r_0)\;.
\end{equation}
Since we are interested in the situation without any matter concentrated on $\Sigma$, we have to ensure that the proper matching conditions \eqref{eq:properconditions} are satisfied (${\mathcal{T}_{\mu\nu}=0}$).

First we focus on the vanishing jump of the extrinsic curvature. The induced metric on $\Sigma$, which coincides from both sides, reads
\begin{equation}
\bs{h}^{-}=\bs{h}^{+}=-p^{\pm}(r_0)\bs{\dd} t^{2}+r_0^{2}(\bs{\dd} \vartheta^{2}+\sin^{2}\vartheta\,\bs{\dd} \varphi^{2})\;.
\end{equation}
Defining the orthonormal 1-forms, which are orthogonal to $\Sigma$ and pointed towards $M^+$,
\begin{equation}
    \bs{n}^{-}=\bs{n}^{+}=\sqrt{q^{\pm}(r_0)}\,\bs{\dd}r\;,
\end{equation}
we obtain the extrinsic curvature from both side of the hypersurface ${r=r_0}$,
\begin{equation}
    \bs{K}^{\pm}=\frac{-p^{\pm}{}'(r_0)}{2\sqrt{q^{\pm}(r_0)}}\bs{\dd}t^2+\frac{r_0}{\sqrt{q^{\pm}(r_0)}}\bs{\dd}\vartheta^2+\frac{r_0\sin^2\vartheta}{\sqrt{q^{\pm}(r_0)}}\bs{\dd}\varphi^2\;.
\end{equation}
The condition ${[\bs{K}]=0}$ together with the continuity of $q$ demands also the continuity of $p'$,
\begin{equation}
    p^{-}{}'(r_0)=p^{+}{}'(r_0)\;.
\end{equation}

Let us move on to the junction conditions involving the Ricci scalar $R$. A significant progress can be made if we suppose that the Ricci scalar is identically zero in the exterior region $M^+$, 
\begin{equation}
    R^+=0\;.
\end{equation}
Under this assumption, the junction conditions for the jump of the Ricci scalar and its derivatives can be rewritten~as 
\begin{equation}
\label{eq:condzeroR}
\left.\Box^{k}R^{-}\right|_{r=r_0}=0\;,
\quad
\left.\partial_{r}\Box^{k}R^{-}\right|_{r=r_0}=0\;,
\quad
k\geq0\;,
\end{equation}
where we used the fact that the Ricci scalar as well as its derivatives depend only on the coordinate $r$. Suppressing ${\pm}$ signs, we write the explicit expression for the action of the wave operator associated with the interior metric on an arbitrary $r$-dependent scalar field ${\phi(r)}$,
\begin{equation}\label{eq:boxphi}
\Box\phi(r)=\alpha\left(r\right)\phi'(r)+\beta\left(r\right)\phi''(r)\;,
\end{equation}
where we denoted
\begin{equation}
\alpha\left(r\right)\equiv\frac{1}{2q(r)^{2}}\left[\left(\frac{p'(r)}{p(r)}+\frac{4}{r}\right)q(r)-q'(r)\right]\;,
\quad 
\beta\left(r\right)\equiv\frac{1}{q(r)}\;.
\end{equation}
Employing \eqref{eq:boxphi}, we obtain the recursive formula for $\Box^k R$,
\begin{equation}
\label{eq:boxRsph}
\Box^{k}R=\alpha\left(r\right)\partial_{r}\Box^{k-1}R+\beta\left(r\right)\partial_{r}^{2}\Box^{k-1}R\;.
\end{equation}

The last expression allows us to rewrite the junction conditions \eqref{eq:condzeroR} purely in terms of partial derivatives with respect to $r$. The reasoning goes as follows: At ${k=0}$, the Ricci scalar and its first derivative must be zero at ${r=r_0}$. At ${k=1}$, the first condition of \eqref{eq:condzeroR} is just a sum of first and second derivatives. Since ${\beta^-(r_0)\neq0}$ (from continuity of $q$ at $r_0$), the second derivative must also vanish. Going to the second condition of \eqref{eq:condzeroR} for ${k=1}$, which is a sum of first, second, and third derivatives, we find that the third derivative is zero as well thanks to ${\beta^-(r_0)\neq0}$. Extending this iterative argument to any $k\geq0$, it is now clear that the conditions \eqref{eq:condzeroR} can be re-expressed as
\begin{equation}\label{eq:rderR}
\left.\partial_{r}^{k}R^{-}\right|_{r=r_0}=0\;,\quad k\geq 0\;.
\end{equation}
Therefore, by assuming analyticity of $R$, we find that ${R}$ vanishes throughout the whole manifold $M$. This gives an explicit equation for $f$ and $g$ that is valid in everywhere in ${M^{-}}$ as well as in ${M^{+}}$,
\begin{equation}\label{eq:ricsczero}
    R=\frac{r^2 q p'^2+r p \left(r p' q'-2 q \left(r p''+2 p'\right)\right)+4 p^2 \left(r q'+q^2-q\right)}{2 r^2 p^2 q^2}=0\;.
\end{equation}
This equation \eqref{eq:ricsczero} and its derivatives enable us to express the $k$-th derivative of functions $p^{-}$ at ${r=r_0}$ as
\begin{equation}
    \pp_r^k p^{-}(r_0) = \Phi_k\big(p^{-}(r_0),p^{-}{}'(r_0),q^{-}(r_0),q^{-}{}'(r_0),\dots, \pp_r^{k-1}q^{-}(r_0)\big)\;,
\end{equation}
where we formally eliminated the dependence on $\pp_r^l p^{-}(r_0)$, ${2\leq l\leq k-1}$, by successive substitutions. The freedom in the choice of $\pp_r^k q^{-}(r_0)$ for ${k\geq 1}$ can be translated to the statement that all the junction conditions are satisfied for an arbitrary function $q^{-}(r)$ and a function $p^{-}(r)$ that is calculated from equation \eqref{eq:ricsczero}.

It is important to stress that all these properties follow purely from the junction conditions. Since the theory is now well-defined in the theory of distributions, we can move on to the field equations. Usually this is an extremely difficult task, but our situation is greatly simplified thanks to the condition \eqref{eq:ricsczero}. Indeed, the field equations of the theory \eqref{eq:fieldeq} reduce everywhere just to
\begin{equation}\label{eq:newfieldeq}
R_{\mu\nu}=T_{\mu\nu}\;,
\end{equation}
which is effectively equivalent to the Einstein field equations with vanishing scalar curvature. For this reason, such solutions with ${R=0}$ everywhere (if they are regular\footnote{A proper treatment of singular geometries in IDG requires non-linear generalized functions.}) are the same as the GR solutions, and thus they are independent of the length scale of non-locality $\mathcal{\ell}$.

Since we are interested in star models with the vacuum outside ${r>r_0}$, we put ${T_{\mu\nu}^+=0}$. This is only solved by the Schwarzschild spacetime,
\begin{equation}
\label{Schwarschild_conditions}
    p^+(r)=1-\frac{2m}{r}\;,
    \quad
    q^+(r)=\bigg(1-\frac{2m}{r}\bigg)^{\!\!-1}\;,
\end{equation}
where $m$ is a positive constant corresponding to mass of the star. The only possible matter allowed in the interior region ${r<r_0}$ (if any regular solution exists) is then given by a traceless energy-momentum tensor ${T^-_{\mu\nu}}$, i.e.,  ${T^-=0}$. As a consequence, the matter inside the star cannot be modelled by common perfect fluids as it usual in GR \cite{bernui1994study}. For example, the well-known Tolman stars \cite{Tolman:1934za} with the Schwarzschild exterior are not solutions of IDG. Thus, if we want to describe the interior of a spherically symmetric and static star with a non-traceless energy-momentum tensor, ${T^-\neq0}$, then the exterior geometry cannot have vanishing Ricci scalar ${R^+}$.


\subsubsection{Collapsing star}

In this section we shall consider a model of a collapsing star where the interior is described by a closed\footnote{The reason we focus on the closed FLRW only is because the open FLRW geometries lead to the scale factors
\begin{equation*}
    a(\tau)=
    \begin{cases}
        \sqrt{(\tau_*-\tau_0)(\tau-\tau_0)}\;, &\epsilon=0\;,
        \\  
        \sqrt{(\tau-\tau_0)^{2}-(\tau_*-\tau_0)^2}\;, &\epsilon<0\;,
    \end{cases}
\end{equation*}
where ${\epsilon}$ denotes the spatial curvature and ${\tau_*}$ and ${\tau_0}$ are constants. These scale factors cannot represent realistic models of collapsing star because they do not have an initial state where the star is at rest as ${\dot{a}(\tau)\neq 0}$.} FLRW metric (positive spatial curvature) in co-moving coordinates,
\begin{equation}
\bs{g}^-=-\bs{\dd} \tau^{2}+a^2\left(\tau\right)\left(\bs{\dd} \lambda^{2}+\sin^{2}\lambda\,(\bs{\dd} \vartheta^{2}+\sin^{2}\vartheta\,\bs{\dd} \varphi^{2})\right)\;,
\end{equation}
where $a\left(\tau\right)$ is the scale factor. The junction surface is located at ${\lambda=\lambda_0}$. The 4-velocity of the observer co-moving with the surface is ${\bs{u}=\bs{\pp}_{\tau}}$, see \cite{poisson2004relativist}. The exterior is given by the spherically symmetric static geometry (with ${p^+>0}$ and ${q^+>0}$),
\begin{equation}
\begin{aligned}
\bs{g}^{+}&=-p^{+}(r)\bs{\dd} t^{2}+q^{+}(r)\bs{\dd} r^{2}+r^{2}\left(\bs{\dd} \vartheta^{2}+\sin^{2}\vartheta\,\bs{\dd} \varphi^{2}\right)\;,
\\
&=-\big(p^{+}(r)\dot{t}^2-q^{+}(r)\dot{r}^2\big)\bs{\dd} \tau^{2}+\big({-}p^{+}(r)t'{}^2+q^{+}(r)r'{}^2\big)\bs{\dd} \lambda^{2}+r^{2}\left(\bs{\dd} \vartheta^{2}+\sin^{2}\vartheta\,\bs{\dd} \varphi^{2}\right)\;,
\end{aligned}
\end{equation}
where we performed a generic transformation ${t=t(\tau,\lambda)}$, ${r=r(\tau,\lambda)}$ in the second line (and hid the explicit dependence on $\tau$ and $\lambda$). We use the shortcuts ${\dot{f}\equiv\pp_{\tau}f}$ and ${f'\equiv\pp_{\lambda}f}$.

First we focus on the restrictions coming from that the continuity of the metric and the continuity of the extrinsic curvature. From the inside the induce metric reads
\begin{equation}\label{eq:hmicol}
\bs{h}^-=-\bs{\dd} \tau^{2}+a^2\left(\tau\right)\sin^{2}\lambda_0\,(\bs{\dd} \vartheta^{2}+\sin^{2}\vartheta\,\bs{\dd} \varphi^{2})\;,
\end{equation}
while from the outside it can be expressed as
\begin{equation}\label{eq:hplcol}
\bs{h}^+=-\Big(p^{+}\big(r(\tau,\lambda_0)\big)\dot{t}^2(\tau,\lambda_0)-q^{+}\big(r(\tau,\lambda_0)\big)\dot{r}^2(\tau,\lambda_0)\Big)\bs{\dd} \tau^{2}+r^2(\tau,\lambda_0)\left(\bs{\dd} \vartheta^{2}+\sin^{2}\vartheta\,\bs{\dd} \varphi^{2}\right)\;.
\end{equation}
By comparing \eqref{eq:hmicol} with \eqref{eq:hplcol}, it becomes clear that the continuity of ${\bs{h}}$ implies
\begin{equation}
r\left(\tau,\lambda_0\right)=a\left(\tau\right)\sin\lambda_{0}
\;,
\quad
\dot{t}(\tau,\lambda_0) p^+(\tau,\lambda_0) =\sqrt{p^+(\tau,\lambda_0)(1+q^+(\tau,\lambda_0)\dot{r}^2(\tau,\lambda_0))}\equiv\Psi(\tau)\;.
\end{equation}
The second expression can be used to obtain $t(\tau,\lambda_0)$ by direct integration
\begin{equation}
    t(\tau,\lambda_0)=\int^{\tau}\!\! d\tilde{\tau} \,\frac{\Psi(\tilde{\tau})}{p^{+}(\tilde{\tau},\lambda_0)}\;.
\end{equation}
Following the arguments in Section~\ref{ssc:matchsurf}, we can now argue that also $\bs{g}$ is continuous. Consider the orthonormal 1-form orthogonal to ${\lambda=\lambda_0}$ and directed towards $M^+$. We immediately see that from the interior it reads
\begin{equation}
\bs{n}^-=a\left(\tau\right)\bs{\dd}\lambda\;.
\end{equation}
Realizing that orthonormal 1-form (pointed towards ${M^+}$) is actually defined uniquely by the conditions ${\bs{n}\cdot\bs{e}_{(a)}=0}$ and ${\bs{n}\cdot\bs{n}^{\sharp}=1}$ with $\bs{e}_{(a)}$ being the vector frame tangent to the hypersurface ${\lambda=\lambda_0}$ (e.g., ${\bs{e}_{(0)}\equiv\bs{u}}$, ${\bs{e}_{(1)}\equiv r^{-1}\bs{\pp}_{\vartheta}}$, ${\bs{e}_{(2)}\equiv (r\sin\vartheta)^{-1}\bs{\pp}_{\varphi}}$), we arrive at
\begin{equation}
    \bs{n}^+=\bs{n}^{-}=-\dot{r}(\tau,\lambda_0)\,\bs{\dd}t+\dot{t}(\tau,\lambda_0)\,\bs{\dd}r\;.
\end{equation}
As a results, the metric $\bs{g}$ is continuous due to ${\bs{g}=\bs{h}+\bs{n}\bs{n}}$. Having defined the orthonormal 1-form, we can now calculate and compare the extrinsic curvatures from both sides of the hypersurface. For the interior side one gets
\begin{equation}
\bs{K}^{-}=a\left(\tau\right)\cos\lambda_0\sin\lambda_0\,(\bs{\dd}\vartheta^2+\sin^{2}\vartheta\bs{\dd}\varphi^2)\;,
\end{equation}
and for the exterior
\begin{equation}
\bs{K}^{+}=\bigg({-}\frac{\dot{\Psi}}{\dot{r}}\bs{\dd} \tau^{2}+a\left(\tau\right)\Psi(\tau)\sin\lambda_0\,(\bs{\dd}\vartheta^2+\sin^{2}\vartheta\bs{\dd}\varphi^2)\bigg)\;.
\end{equation}
By demanding that the jump of the extrinsic curvature is zero, ${\bs{K}}$, we find that $\Psi(\tau)$ is constant,
\begin{equation}
\Psi(\tau)=\cos\lambda_{0}\;.
\end{equation}
Hence, from the continuity of the metric and the extrinsic curvature have obtained the following relations
\begin{equation}
    r\left(\tau,\lambda_0\right)=a\left(\tau\right)\sin\lambda_{0}\;,
    \quad
     t(\tau,\lambda_0)=\cos\lambda_{0}\int^{\tau}\!\!  \,\frac{d\tilde{\tau}}{p^{+}(\tilde{\tau},\lambda_0)}\;.
\end{equation}

Let us now focus on the remaining junction conditions, which put restrictions on the jump of the Ricci scalar. Assuming that the exterior geometry has vanishing Ricci scalar, ${R^+=0}$, we can write in co-moving coordinates
\begin{equation}\label{eq:juncFLRWzeroR}
\Box^{k}R^{-}\big|_{\lambda=\lambda_{0}}=0\;,
\qquad
\partial_{\tau}\Box^{k}R^{-}\big|_{\lambda=\lambda_{0}}=0\;,\qquad k\geq0\;.
\end{equation}
The Ricci scalar of the interior closed FLRW metric is given by
\begin{equation}
\label{ricciFLRW}
R^{-}=\frac{6}{a^{2}}\left(1+\dot{a}^{2}+a\ddot{a}\right)\;.
\end{equation}
Since the repeated action of the wave operator reads
\begin{equation}
\Box^{k}R^{-}=-3\frac{\dot{a}\left(\tau\right)}{a\left(\tau\right)}\partial_{\tau}\Box^{k-1}R^{-}-\partial_{\tau}^{2}\Box^{k-1}R^{-}\;,
\end{equation}
We can make a very similar argument to the one we did for the static case. Following these steps, one can find that the conditions \eqref{eq:juncFLRWzeroR} can be rewritten as
\begin{equation}
\partial_{\tau}^{k}R^{-}\big|_{\lambda=\lambda_0}=0\;,\qquad k\geq0\;.
\end{equation}
Again, assuming the analyticity of $R$, this clearly implies that the Ricci scalar must be zero not only in the exterior but also throughout the whole interior region. The equation ${R^{-}=0}$ has the following solution
\begin{equation}\label{eq:scalefactor}
a\left(\tau\right)=\sqrt{(\tau_*-\tau_0)^2-(\tau-\tau_0)^{2}}\;,
\end{equation}
where $\tau_*$ and $\tau_0$ are arbitrary constants. As before, we can see that the non-local character of the theory causes that the constraints themselves affect the whole spacetime and determine the geometry to a great extent. In particular, they imply that the Ricci scalar must vanish everywhere, ${R=0}$. Thus, even without solving the field equations we already see that the our simple model with the closed FLRW interior and spherically symmetric static exterior of vanishing Ricci scalar (e.g., the Schwarzschild metric) does not allow bounce. A more complicated geometry in the exterior is required.

Finally let us comment on the interior geometry with \eqref{eq:scalefactor}. The scale-factor ${a(\tau)}$ has the maximum value at ${\tau=\tau_0}$ and decreases towards ${\tau=\tau_*}$, which corresponds to the curvature singularity as can be seen from the Kretschmann scalar,
\begin{equation}
    R_{\mu\nu\kappa\lambda}R^{\mu\nu\kappa\lambda}=\frac{24 (\tau_*-\tau_0)^4}{(\tau-\tau_*)^4 (\tau-2 \tau_0+\tau_*)^4}\;.
\end{equation}
Unfortunately, due to the singular behaviour the check whether this geometry is a regular solution is highly non-trivial and goes beyond the scope of this paper. Our statement that ${R=0}$ is only valid outside the point of singularity ${\tau<\tau_*}$. In order to check that a singular geometry is a solution of IDG one should first include the singular point to the manifold as well and describe the Ricci scalar using non-linear generalized functions localized at ${\tau=\tau_*}$.\footnote{See \cite{Heinzle:2001bk}, for such a description of the singularity in the Schwarzschild geometry.} As a next step one could calculate the action of $F(\Box)$ on $R$, for instance, using the decomposition into the eigenfunctions of $\Box$. Such a study is very challenging and has never been done even for the simplest cases such as the Schwarzschild metric.

\section{Conclusions}\label{sc:conclusion}

In this paper we found the junction conditions, junction field equations for the non-local infinite derivative theory ${R+RF(\Box)R}$ (IDG). To achieve that, we relied on the theory of linear distributions to describe the potential discontinuities and we assumed that the conditions can be imposed by avoiding ill-defined expressions term by term in the infinite sums (ignoring their regularizing and deregularizing properties). After introducing necessary tools of distributional theory and the infinite derivative gravity, we obtained the matching conditions \eqref{eq:conditions} by requiring that the theory is well-behaved in a distributional sense, These conditions are much more restrictive than in the general relativity (GR) or other local theories of gravity. They say that the trace of the extrinsic curvature as well as ${\Box^k R}$ and ${\bs{\nabla}\Box^k R}$ for all ${k\geq 0}$ must be continuous across the matching hypersurface.

The junction field equations and the allowed matter content on the hypersurface get also modified in IDG, see \eqref{eq:juncfieeq} with \eqref{eq:tautautau}. Similar to GR, the external flux and external tension are not present (which is not the case of ${R+R^2}$, for example), but unlike in GR the energy-momentum on the hypersurface must be traceless. From there it also follows that the proper matching with no matter on the hypersurface requires vanishing jump of the extrinsic curvature. 

Finally, we illustrated the use of our results in several physically motivated scenarios. First, we focused on the implications for the braneworld models with 5-dimensional bulk theory being IDG. It turns out that the internal tension of the brane is then given purely by the trace of the energy-momentum tensor of the matter confined to the brane. Next, we studied the consequences of the junction conditions on two simple examples of static and collapsing stars. In particular, we found that having a spherically symmetric static exterior with zero Ricci scalar implies that the Ricci scalar in the interior region must either be non-analytic or also vanish. These examples demonstrate that the junction condition of non-local theories hugely constraint the possible metrics that one can consider even before going into the field equations. Of course, much more difficult task is to check whether the resulting geometry actually is a solution or not. To answer this question, new methods of solving non-local nonlinear equations in curved geometry must be developed.

To conclude, let us point out that this work can be extended by considering terms of the form $R_{\mu\nu}F_2(\Box)R^{\mu\nu}$ and $R_{\mu\nu\kappa\lambda}F_3(\Box)R^{\mu\nu\kappa\lambda}$ in the action. This will of course impose more constraints on the jumps of the curvature, possibly reaching a point where only completely smooth metrics with no matter content on the hypersurface will be allowed. An even more interesting problem is the study that goes beyond our simplifying assumptions especially the case where the non-local operator can produce a singular part when it acts on a smooth function. We plan to investigate these extensions in our future works.


\section*{Acknowledgements}

IK  and AM are supported by Netherlands Organization for Scientific Research (NWO) Grant No.680-91-119. FJMT acknowledges financial support from NRF Grants No.120390, Reference: BSFP190416431035; No.120396, Reference: CSRP190405427545; No.101775, Reference: SFH150727131568; and the financial support from the NASSP Programme - UCT node and the Van Swinderen Institute at the University of Groningen.


\bibliographystyle{JHEP}
\bibliography{references}

\providecommand{\href}[2]{#2}\begingroup\raggedright\begin{thebibliography}{10}

\bibitem{Misner:1974qy}
C.~W. Misner, K.~Thorne and J.~Wheeler, \emph{{Gravitation}}.
\newblock W. H. Freeman, San Francisco, 1973.

\bibitem{https://doi.org/10.1002/andp.19243791403}
K.~Lanczos, \emph{Flächenhafte verteilung der materie in der einsteinschen
  gravitationstheorie},
  \href{http://dx.doi.org/https://doi.org/10.1002/andp.19243791403}{\emph{Annalen
  der Physik} {\bfseries 379} (1924) 518--540},
  [\href{https://arxiv.org/abs/https://onlinelibrary.wiley.com/doi/pdf/10.1002/andp.19243791403}{{\ttfamily
  https://onlinelibrary.wiley.com/doi/pdf/10.1002/andp.19243791403}}].

\bibitem{darmois1927equations}
G.~Darmois, \emph{Les {\'e}quations de la gravitation einsteinienne}.
\newblock Gauthier-Villars Paris, 1927.

\bibitem{Lichnerowicz:107002}
A.~Lichnerowicz, \emph{{Théories relativistes de la gravitation et de
  l'électromagnétisme: relativité générale et théories unitaires}}.
\newblock Masson, Paris, 1955.

\bibitem{o1952jump}
S.~O'Brien and J.~L. Synge, \emph{Jump conditions at discontinuities in general
  relativity}, {\emph{Commun. Dublin Inst. Adv. Stud., A} (1952) }.

\bibitem{bel1967conditions}
L.~Bel and A.~Hamoui, \emph{Les conditions de raccordement en relativit{\'e}
  g{\'e}n{\'e}rale},  in \emph{Annales de l'IHP Physique th{\'e}orique},
  vol.~7, pp.~229--244, 1967.

\bibitem{taub1980space}
A.~Taub, \emph{Spacetimes with distribution valued curvature tensors},
  {\emph{Journal of Mathematical Physics} {\bfseries 21} (1980) 1423--1431}.

\bibitem{bonnor1981junction}
W.~Bonnor and P.~Vickers, \emph{Junction conditions in general relativity},
  {\emph{General Relativity and Gravitation} {\bfseries 13} (1981) 29--36}.

\bibitem{clarke1987junction}
C.~Clarke and T.~Dray, \emph{Junction conditions for null hypersurfaces},
  {\emph{Classical and Quantum Gravity} {\bfseries 4} (1987) 265}.

\bibitem{barrabes1989singular}
C.~Barrabes, \emph{Singular hypersurfaces in general relativity: a unified
  description}, {\emph{Classical and Quantum Gravity} {\bfseries 6} (1989)
  581}.

\bibitem{Mars:1993mj}
M.~Mars and J.~M. Senovilla, \emph{{Geometry of general hypersurfaces in
  space-time: Junction conditions}},
  \href{http://dx.doi.org/10.1088/0264-9381/10/9/026}{\emph{Class. Quant.
  Grav.} {\bfseries 10} (1993) 1865--1897},
  [\href{https://arxiv.org/abs/gr-qc/0201054}{{\ttfamily gr-qc/0201054}}].

\bibitem{Israel:1966rt}
W.~Israel, \emph{{Singular hypersurfaces and thin shells in general
  relativity}}, \href{http://dx.doi.org/10.1007/BF02710419}{\emph{Nuovo Cim. B}
  {\bfseries 44S10} (1966) 1}.

\bibitem{barrabes1991thin}
C.~Barrabes and W.~Israel, \emph{Thin shells in general relativity and
  cosmology: The lightlike limit}, {\emph{Physical Review D} {\bfseries 43}
  (1991) 1129}.

\bibitem{Davis:2002gn}
S.~C. Davis, \emph{{Generalized Israel junction conditions for a Gauss-Bonnet
  brane world}},
  \href{http://dx.doi.org/10.1103/PhysRevD.67.024030}{\emph{Phys. Rev. D}
  {\bfseries 67} (2003) 024030},
  [\href{https://arxiv.org/abs/hep-th/0208205}{{\ttfamily hep-th/0208205}}].

\bibitem{Battye:2001pb}
R.~A. Battye and B.~Carter, \emph{{Generic junction conditions in brane world
  scenarios}},
  \href{http://dx.doi.org/10.1016/S0370-2693(01)00495-6}{\emph{Phys. Lett. B}
  {\bfseries 509} (2001) 331--336},
  [\href{https://arxiv.org/abs/hep-th/0101061}{{\ttfamily hep-th/0101061}}].

\bibitem{Deruelle:2007pt}
N.~Deruelle, M.~Sasaki and Y.~Sendouda, \emph{{Junction conditions in f(R)
  theories of gravity}},
  \href{http://dx.doi.org/10.1143/PTP.119.237}{\emph{Prog. Theor. Phys.}
  {\bfseries 119} (2008) 237--251},
  [\href{https://arxiv.org/abs/0711.1150}{{\ttfamily 0711.1150}}].

\bibitem{Clifton:2012ry}
T.~Clifton, P.~Dunsby, R.~Goswami and A.~M. Nzioki, \emph{{On the absence of
  the usual weak-field limit, and the impossibility of embedding some known
  solutions for isolated masses in cosmologies with f(R) dark energy}},
  \href{http://dx.doi.org/10.1103/PhysRevD.87.063517}{\emph{Phys. Rev. D}
  {\bfseries 87} (2013) 063517},
  [\href{https://arxiv.org/abs/1210.0730}{{\ttfamily 1210.0730}}].

\bibitem{Senovilla:2013vra}
J.~M. Senovilla, \emph{{Junction conditions for F(R)-gravity and their
  consequences}},
  \href{http://dx.doi.org/10.1103/PhysRevD.88.064015}{\emph{Phys. Rev. D}
  {\bfseries 88} (2013) 064015},
  [\href{https://arxiv.org/abs/1303.1408}{{\ttfamily 1303.1408}}].

\bibitem{Olmo:2020fri}
G.~J. Olmo and D.~Rubiera-Garcia, \emph{{Junction conditions in Palatini $f(R)$
  gravity}}, \href{http://dx.doi.org/10.1088/1361-6382/abb924}{\emph{Class.
  Quant. Grav.} {\bfseries 37} (2020) 215002},
  [\href{https://arxiv.org/abs/2007.04065}{{\ttfamily 2007.04065}}].

\bibitem{Vignolo:2018eco}
S.~Vignolo, R.~Cianci and S.~Carloni, \emph{{On the junction conditions in
  $f(R)$-gravity with torsion}},
  \href{http://dx.doi.org/10.1088/1361-6382/aab6fe}{\emph{Class. Quant. Grav.}
  {\bfseries 35} (2018) 095014},
  [\href{https://arxiv.org/abs/1801.08344}{{\ttfamily 1801.08344}}].

\bibitem{Reina:2015gxa}
B.~Reina, J.~M.~M. Senovilla and R.~Vera, \emph{{Junction conditions in
  quadratic gravity: thin shells and double layers}},
  \href{http://dx.doi.org/10.1088/0264-9381/33/10/105008}{\emph{Class. Quant.
  Grav.} {\bfseries 33} (2016) 105008},
  [\href{https://arxiv.org/abs/1510.05515}{{\ttfamily 1510.05515}}].

\bibitem{Arkuszewski:1975fz}
W.~Arkuszewski, W.~Kopczynski and V.~Ponomarev, \emph{{Matching Conditions in
  the Einstein-Cartan Theory of Gravitation}},
  \href{http://dx.doi.org/10.1007/BF01629248}{\emph{Commun. Math. Phys.}
  {\bfseries 45} (1975) 183--190}.

\bibitem{Macias:2002sr}
A.~Macias, C.~Lammerzahl and L.~O. Pimentel, \emph{{Matching conditions in
  metric affine gravity}},
  \href{http://dx.doi.org/10.1103/PhysRevD.66.104013}{\emph{Phys. Rev. D}
  {\bfseries 66} (2002) 104013}.

\bibitem{Padilla:2012ze}
A.~Padilla and V.~Sivanesan, \emph{{Boundary Terms and Junction Conditions for
  Generalized Scalar-Tensor Theories}},
  \href{http://dx.doi.org/10.1007/JHEP08(2012)122}{\emph{JHEP} {\bfseries 08}
  (2012) 122}, [\href{https://arxiv.org/abs/1206.1258}{{\ttfamily 1206.1258}}].

\bibitem{delaCruz-Dombriz:2014zaa}
A.~de~la Cruz-Dombriz, P.~K.~S. Dunsby and D.~Saez-Gomez, \emph{{Junction
  conditions in extended Teleparallel gravities}},
  \href{http://dx.doi.org/10.1088/1475-7516/2014/12/048}{\emph{JCAP} {\bfseries
  12} (2014) 048}, [\href{https://arxiv.org/abs/1406.2334}{{\ttfamily
  1406.2334}}].

\bibitem{Biswas:2005qr}
T.~Biswas, A.~Mazumdar and W.~Siegel, \emph{{Bouncing universes in
  string-inspired gravity}},
  \href{http://dx.doi.org/10.1088/1475-7516/2006/03/009}{\emph{JCAP} {\bfseries
  03} (2006) 009}, [\href{https://arxiv.org/abs/hep-th/0508194}{{\ttfamily
  hep-th/0508194}}].

\bibitem{Biswas:2011ar}
T.~Biswas, E.~Gerwick, T.~Koivisto and A.~Mazumdar, \emph{{Towards singularity
  and ghost free theories of gravity}},
  \href{http://dx.doi.org/10.1103/PhysRevLett.108.031101}{\emph{Phys. Rev.
  Lett.} {\bfseries 108} (2012) 031101},
  [\href{https://arxiv.org/abs/1110.5249}{{\ttfamily 1110.5249}}].

\bibitem{Biswas:2016etb}
T.~Biswas, A.~S. Koshelev and A.~Mazumdar, \emph{{Gravitational theories with
  stable (anti-)de Sitter backgrounds}},
  \href{http://dx.doi.org/10.1007/978-3-319-31299-6_5}{\emph{Fundam. Theor.
  Phys.} {\bfseries 183} (2016) 97--114},
  [\href{https://arxiv.org/abs/1602.08475}{{\ttfamily 1602.08475}}].

\bibitem{Biswas:2010zk}
T.~Biswas, T.~Koivisto and A.~Mazumdar, \emph{{Towards a resolution of the
  cosmological singularity in non-local higher derivative theories of
  gravity}}, \href{http://dx.doi.org/10.1088/1475-7516/2010/11/008}{\emph{JCAP}
  {\bfseries 11} (2010) 008},
  [\href{https://arxiv.org/abs/1005.0590}{{\ttfamily 1005.0590}}].

\bibitem{tomboulis1997superrenormalizable}
E.~Tomboulis, \emph{Superrenormalizable gauge and gravitational theories},
  {\emph{arXiv preprint hep-th/9702146} (1997) }.

\bibitem{Modesto:2011kw}
L.~Modesto, \emph{{Super-renormalizable Quantum Gravity}},
  \href{http://dx.doi.org/10.1103/PhysRevD.86.044005}{\emph{Phys. Rev. D}
  {\bfseries 86} (2012) 044005},
  [\href{https://arxiv.org/abs/1107.2403}{{\ttfamily 1107.2403}}].

\bibitem{Talaganis:2014ida}
S.~Talaganis, T.~Biswas and A.~Mazumdar, \emph{{Towards understanding the
  ultraviolet behavior of quantum loops in infinite-derivative theories of
  gravity}},
  \href{http://dx.doi.org/10.1088/0264-9381/32/21/215017}{\emph{Class. Quant.
  Grav.} {\bfseries 32} (2015) 215017},
  [\href{https://arxiv.org/abs/1412.3467}{{\ttfamily 1412.3467}}].

\bibitem{Abel:2019ufz}
S.~Abel and N.~A. Dondi, \emph{{UV Completion on the Worldline}},
  \href{http://dx.doi.org/10.1007/JHEP07(2019)090}{\emph{JHEP} {\bfseries 07}
  (2019) 090}, [\href{https://arxiv.org/abs/1905.04258}{{\ttfamily
  1905.04258}}].

\bibitem{Abel:2019zou}
S.~Abel, L.~Buoninfante and A.~Mazumdar, \emph{{Nonlocal gravity with worldline
  inversion symmetry}},
  \href{http://dx.doi.org/10.1007/JHEP01(2020)003}{\emph{JHEP} {\bfseries 01}
  (2020) 003}, [\href{https://arxiv.org/abs/1911.06697}{{\ttfamily
  1911.06697}}].

\bibitem{Abel:2020gdi}
S.~Abel and D.~Lewis, \emph{{Worldline theories with towers of internal
  states}}, \href{http://dx.doi.org/10.1007/JHEP12(2020)069}{\emph{JHEP}
  {\bfseries 12} (2020) 069},
  [\href{https://arxiv.org/abs/2007.07242}{{\ttfamily 2007.07242}}].

\bibitem{Calcagni:2018lyd}
G.~Calcagni, L.~Modesto and G.~Nardelli, \emph{{Initial conditions and degrees
  of freedom of non-local gravity}},
  \href{http://dx.doi.org/10.1007/JHEP05(2018)087}{\emph{JHEP} {\bfseries 05}
  (2018) 087}, [\href{https://arxiv.org/abs/1803.00561}{{\ttfamily
  1803.00561}}].

\bibitem{Kolar:2020ezu}
I.~Kolar and A.~Mazumdar, \emph{{Hamiltonian for scalar field model of infinite
  derivative gravity}},
  \href{http://dx.doi.org/10.1103/PhysRevD.101.124028}{\emph{Phys. Rev. D}
  {\bfseries 101} (2020) 124028},
  [\href{https://arxiv.org/abs/2003.00590}{{\ttfamily 2003.00590}}].

\bibitem{Buoninfante:2018stt}
L.~Buoninfante, G.~Harmsen, S.~Maheshwari and A.~Mazumdar, \emph{{Nonsingular
  metric for an electrically charged point-source in ghost-free infinite
  derivative gravity}},
  \href{http://dx.doi.org/10.1103/PhysRevD.98.084009}{\emph{Phys. Rev. D}
  {\bfseries 98} (2018) 084009},
  [\href{https://arxiv.org/abs/1804.09624}{{\ttfamily 1804.09624}}].

\bibitem{Boos:2018bxf}
J.~Boos, V.~P. Frolov and A.~Zelnikov, \emph{{Gravitational field of static p
  -branes in linearized ghost-free gravity}},
  \href{http://dx.doi.org/10.1103/PhysRevD.97.084021}{\emph{Phys. Rev. D}
  {\bfseries 97} (2018) 084021},
  [\href{https://arxiv.org/abs/1802.09573}{{\ttfamily 1802.09573}}].

\bibitem{Boos:2020ccj}
J.~Boos, J.~Pinedo~Soto and V.~P. Frolov, \emph{{Ultrarelativistic spinning
  objects in nonlocal ghost-free gravity}},
  \href{http://dx.doi.org/10.1103/PhysRevD.101.124065}{\emph{Phys. Rev. D}
  {\bfseries 101} (2020) 124065},
  [\href{https://arxiv.org/abs/2004.07420}{{\ttfamily 2004.07420}}].

\bibitem{Kolar:2020bpo}
I.~Kol\'a\v{r} and A.~Mazumdar, \emph{{NUT charge in linearized infinite
  derivative gravity}},
  \href{http://dx.doi.org/10.1103/PhysRevD.101.124005}{\emph{Phys. Rev. D}
  {\bfseries 101} (2020) 124005},
  [\href{https://arxiv.org/abs/2004.07613}{{\ttfamily 2004.07613}}].

\bibitem{Buoninfante:2018xif}
L.~Buoninfante, A.~S. Cornell, G.~Harmsen, A.~S. Koshelev, G.~Lambiase, J.~a.
  Marto et~al., \emph{{Towards nonsingular rotating compact object in
  ghost-free infinite derivative gravity}},
  \href{http://dx.doi.org/10.1103/PhysRevD.98.084041}{\emph{Phys. Rev. D}
  {\bfseries 98} (2018) 084041},
  [\href{https://arxiv.org/abs/1807.08896}{{\ttfamily 1807.08896}}].

\bibitem{Frolov:2015usa}
V.~P. Frolov and A.~Zelnikov, \emph{{Head-on collision of ultrarelativistic
  particles in ghost-free theories of gravity}},
  \href{http://dx.doi.org/10.1103/PhysRevD.93.064048}{\emph{Phys. Rev. D}
  {\bfseries 93} (2016) 064048},
  [\href{https://arxiv.org/abs/1509.03336}{{\ttfamily 1509.03336}}].

\bibitem{Frolov:2015bia}
V.~P. Frolov, A.~Zelnikov and T.~de~Paula~Netto, \emph{{Spherical collapse of
  small masses in the ghost-free gravity}},
  \href{http://dx.doi.org/10.1007/JHEP06(2015)107}{\emph{JHEP} {\bfseries 06}
  (2015) 107}, [\href{https://arxiv.org/abs/1504.00412}{{\ttfamily
  1504.00412}}].

\bibitem{Frolov:2015bta}
V.~P. Frolov, \emph{{Mass-gap for black hole formation in higher derivative and
  ghost free gravity}},
  \href{http://dx.doi.org/10.1103/PhysRevLett.115.051102}{\emph{Phys. Rev.
  Lett.} {\bfseries 115} (2015) 051102},
  [\href{https://arxiv.org/abs/1505.00492}{{\ttfamily 1505.00492}}].

\bibitem{Kilicarslan:2018yxd}
E.~Kilicarslan, \emph{{Weak Field Limit of Infinite Derivative Gravity}},
  \href{http://dx.doi.org/10.1103/PhysRevD.98.064048}{\emph{Phys. Rev. D}
  {\bfseries 98} (2018) 064048},
  [\href{https://arxiv.org/abs/1808.00266}{{\ttfamily 1808.00266}}].

\bibitem{Dengiz:2020xbu}
S.~Dengiz, E.~Kilicarslan, I.~Kol\'a\v{r} and A.~Mazumdar, \emph{{Impulsive
  waves in ghost free infinite derivative gravity in anti-de Sitter
  spacetime}}, \href{http://dx.doi.org/10.1103/PhysRevD.102.044016}{\emph{Phys.
  Rev. D} {\bfseries 102} (2020) 044016},
  [\href{https://arxiv.org/abs/2006.07650}{{\ttfamily 2006.07650}}].

\bibitem{delaCruz-Dombriz:2018aal}
A.~de~la Cruz-Dombriz, F.~J. Maldonado~Torralba and A.~Mazumdar,
  \emph{{Nonsingular and ghost-free infinite derivative gravity with torsion}},
  \href{http://dx.doi.org/10.1103/PhysRevD.99.104021}{\emph{Phys. Rev. D}
  {\bfseries 99} (2019) 104021},
  [\href{https://arxiv.org/abs/1812.04037}{{\ttfamily 1812.04037}}].

\bibitem{de2019ghost}
{\'A}.~de~la Cruz-Dombriz, F.~J.~M. Torralba and A.~Mazumdar, \emph{Ghost-free
  higher-order theories of gravity with torsion}, {\emph{arXiv preprint
  arXiv:1911.08846} (2019) }.

\bibitem{Schwartz}
L.~Schwartz, \emph{Sur l'impossibilit\'e de la multiplication des
  distributions}, {\emph{C. R. Acad. Sci. Paris} {\bfseries 239} (1954)
  847--8}.

\bibitem{colombeau1990}
J.~F. Colombeau, \emph{Multiplication of distributions}, {\emph{Bull. Amer.
  Math. Soc. (N.S.)} {\bfseries 23} (10, 1990) 251--268}.

\bibitem{Steinbauer:2006qi}
R.~Steinbauer and J.~A. Vickers, \emph{{The Use of generalised functions and
  distributions in general relativity}},
  \href{http://dx.doi.org/10.1088/0264-9381/23/10/R01}{\emph{Class. Quant.
  Grav.} {\bfseries 23} (2006) R91--R114},
  [\href{https://arxiv.org/abs/gr-qc/0603078}{{\ttfamily gr-qc/0603078}}].

\bibitem{Grosser:1620651}
M.~Grosser, M.~Kunzinger, M.~Oberguggenberger and R.~Steinbauer,
  \emph{{Geometric theory of generalized functions with applications to general
  relativity}}.
\newblock Mathematics and Its Applications. Springer, Dordrecht, 2001,
  \href{http://dx.doi.org/10.1007/978-94-015-9845-3}{10.1007/978-94-015-9845-3}.

\bibitem{Biswas:2013cha}
T.~Biswas, A.~Conroy, A.~S. Koshelev and A.~Mazumdar, \emph{{Generalized
  ghost-free quadratic curvature gravity}},
  \href{http://dx.doi.org/10.1088/0264-9381/31/1/015022}{\emph{Class. Quant.
  Grav.} {\bfseries 31} (2014) 015022},
  [\href{https://arxiv.org/abs/1308.2319}{{\ttfamily 1308.2319}}].

\bibitem{Senovilla:2014kua}
J.~M. Senovilla, \emph{{Gravitational double layers}},
  \href{http://dx.doi.org/10.1088/0264-9381/31/7/072002}{\emph{Class. Quant.
  Grav.} {\bfseries 31} (2014) 072002},
  [\href{https://arxiv.org/abs/1402.1139}{{\ttfamily 1402.1139}}].

\bibitem{Maartens:2010ar}
R.~Maartens and K.~Koyama, \emph{{Brane-World Gravity}},
  \href{http://dx.doi.org/10.12942/lrr-2010-5}{\emph{Living Rev. Rel.}
  {\bfseries 13} (2010) 5}, [\href{https://arxiv.org/abs/1004.3962}{{\ttfamily
  1004.3962}}].

\bibitem{Flanagan:2000nx}
E.~Flanagan, N.~T. Jones, H.~Stoica, S.~Tye and I.~Wasserman, \emph{{A Brane
  world perspective on the cosmological constant and the hierarchy problems}},
  \href{http://dx.doi.org/10.1103/PhysRevD.64.045007}{\emph{Phys. Rev. D}
  {\bfseries 64} (2001) 045007},
  [\href{https://arxiv.org/abs/hep-th/0012129}{{\ttfamily hep-th/0012129}}].

\bibitem{Randall:1999ee}
L.~Randall and R.~Sundrum, \emph{{A Large mass hierarchy from a small extra
  dimension}}, \href{http://dx.doi.org/10.1103/PhysRevLett.83.3370}{\emph{Phys.
  Rev. Lett.} {\bfseries 83} (1999) 3370--3373},
  [\href{https://arxiv.org/abs/hep-ph/9905221}{{\ttfamily hep-ph/9905221}}].

\bibitem{Randall:1999vf}
L.~Randall and R.~Sundrum, \emph{{An Alternative to compactification}},
  \href{http://dx.doi.org/10.1103/PhysRevLett.83.4690}{\emph{Phys. Rev. Lett.}
  {\bfseries 83} (1999) 4690--4693},
  [\href{https://arxiv.org/abs/hep-th/9906064}{{\ttfamily hep-th/9906064}}].

\bibitem{Shiromizu:1999wj}
T.~Shiromizu, K.-i. Maeda and M.~Sasaki, \emph{{The Einstein equation on the
  3-brane world}},
  \href{http://dx.doi.org/10.1103/PhysRevD.62.024012}{\emph{Phys. Rev. D}
  {\bfseries 62} (2000) 024012},
  [\href{https://arxiv.org/abs/gr-qc/9910076}{{\ttfamily gr-qc/9910076}}].

\bibitem{Forste:2000ge}
S.~Forste, \emph{{Fine tuning of the cosmological constant in brane worlds}},
  \href{http://dx.doi.org/10.1002/1521-3978(200105)49:4/6<495::AID-PROP495>3.0.CO;2-Z}{\emph{Fortsch.
  Phys.} {\bfseries 49} (2001) 495--501},
  [\href{https://arxiv.org/abs/hep-th/0012029}{{\ttfamily hep-th/0012029}}].

\bibitem{hawking_ellis_1973}
S.~W. Hawking and G.~F.~R. Ellis, \emph{The Large Scale Structure of
  Space-Time}.
\newblock Cambridge Monographs on Mathematical Physics. Cambridge University
  Press, 1973,
  \href{http://dx.doi.org/10.1017/CBO9780511524646}{10.1017/CBO9780511524646}.

\bibitem{bernui1994study}
A.~Bernui and E.~Portocarrero, \emph{Study of the matching of solutions of the
  {Einstein} equations according to {Darmois}}, {\emph{The Astrophysical
  Journal} {\bfseries 427} (1994) 947--950}.

\bibitem{Tolman:1934za}
R.~C. Tolman, \emph{{Effect of imhomogeneity on cosmological models}},
  \href{http://dx.doi.org/10.1073/pnas.20.3.169}{\emph{Proc. Nat. Acad. Sci.}
  {\bfseries 20} (1934) 169--176}.

\bibitem{poisson2004relativist}
E.~Poisson, \emph{A relativist's toolkit: the mathematics of black-hole
  mechanics}.
\newblock Cambridge university press, 2004.

\bibitem{Heinzle:2001bk}
J.~Heinzle and R.~Steinbauer, \emph{{Remarks on the distributional
  Schwarzschild geometry}}, \href{http://dx.doi.org/10.1063/1.1448684}{\emph{J.
  Math. Phys.} {\bfseries 43} (2002) 1493--1508},
  [\href{https://arxiv.org/abs/gr-qc/0112047}{{\ttfamily gr-qc/0112047}}].

\end{thebibliography}\endgroup

\end{document}